\documentclass[12pt,onecolumn]{IEEEtran}

\usepackage{latexsym}
\usepackage{amsfonts}
\usepackage{amsbsy}
\usepackage{graphicx}
\usepackage{subfigure}
\usepackage{float}
\usepackage{algorithm}
\usepackage{algorithmic}
\usepackage{flushend}
\usepackage{mathrsfs}
\usepackage{amsfonts}
\usepackage{array}
\usepackage{amssymb,amsmath}
\usepackage{cases}
\usepackage{longtable}
\usepackage{rotating}
\usepackage{multirow}
\usepackage{epstopdf}
\usepackage{flushend}
\usepackage{cite}
\usepackage{float}
\usepackage[marginal]{footmisc}
\usepackage{stfloats}
\usepackage[colorlinks,
            linkcolor=black,
            anchorcolor=black,
             urlcolor=blue,
            citecolor=black]{hyperref}

\def\BibTeX{{\rm B\kern-.05em{\sc i\kern-.025em b}\kern-.08em
    T\kern-.1667em\lower.7ex\hbox{E}\kern-.125emX}}

\newtheorem{proposition}{Proposition}

\newtheorem{theorem}{Theorem}

\newtheorem{lemma}{Lemma}

\begin{document}

\title{ UAV-Enabled Cooperative Jamming for Covert Communications }

\author{Hangmei Rao,
         Sa Xiao,
         Shihao Yan,~\IEEEmembership{Member,~IEEE},
         Jianquan Wang,

         and
         Wanbin Tang,~\IEEEmembership{Member,~IEEE}

\thanks{
H. Rao, S.Xiao, J. Wang and W. Tang  are with the National Key Laboratory of Science
and Technology on Communications, University of Electronic
Science and Technology of China, Chengdu 611731, China (Email: yqr@std.uestc.edu.cn; xiaosa@uestc.edu.cn; wjq2002wjq@126.com; wbtang@uestc.edu.cn)}.
\thanks{S. Yan is with the School of Electrical Engineering and Telecommunications, the University of New South Wales, Sydney, NSW 2052, Australia (Email: shihao.yan@unsw.edu.au).
}
}

\markboth{~~}%
{Shell \MakeLowercase{\textit{et al.}}: }

\maketitle
\begin{abstract}
This work employs an unmanned aerial vehicle (UAV) as a jammer to aid a covert communication from a transmitter Alice to a receiver Bob, where the UAV transmits artificial noise (AN) with random power to deliberately create interference to a warden Willie. In the considered system, the UAV's trajectory is critical to the covert communication performance, since the AN transmitted by the UAV also generates interference to Bob. To maximize the system performance, we formulate an optimization problem to jointly design the UAV's trajectory and Alice's transmit power. The formulated optimization problem is non-convex and is normally solved by a conventional iterative (CI) method, which requires multiple approximations based on Taylor expansions and an initialization on the UAV's trajectory. In order to eliminate these requirements, this work, for the first time, develops a geometric (GM) method  to solve the optimization problem. By analyzing the covertness constraint, the GM method decouples the joint optimization into optimizing the UAV's trajectory and Alice's transmit power separately. Our examination shows that the GM method can significantly outperform the CI method in terms of achieving a higher average covert rate and the complexity of the GM method is lower than that of the CI method.
\end{abstract}


\IEEEpeerreviewmaketitle

\section{Introduction}

Due to the broadcast nature of the wireless medium, both legitimate and illegitimate transceivers can receive data signals from the wireless air interface. Therefore, how to guarantee communication security and privacy is of a paramount significance. Against this background, in the context of physical layer security various strategies have been proposed to prevent the information being eavesdropped on by an eavesdropper~\cite{Wyner,Cachin,Sheikholeslami1,Sheikholeslami2}. However, in many practical scenarios, such as military applications, the existence of a wireless transmission is not allowed to be discovered by a third transceiver in order to preserve a user's privacy. To avoid the data transmission from being discovered, covert or low probability of detection (LPD) communications have been proposed and studied in both industry and academia.
In covert communications, a transmitter Alice intends to transmit data signals to a receiver Bob covertly, while a warden Willie tries to detect Alice's transmission. If the detection performance of Willie is close to that of a random guess, the transmission from Alice to Bob can be regarded as covert.

In the literature, there have been multiple studies aiming at analyzing the achievable rate of covert communications from the perspective of information theory. The covert rate was firstly analyzed over additive white Gaussian noise (AWGN) channels in \cite{Bash2}. Then, the result was extended to discrete memoryless channels (DMC) in \cite{covert_resolv} and \cite{Wang1}, binary symmetric channels (BSC) in \cite{Che[]}, and multiple access channels (MAC) in \cite{Arumugam}.
The aforementioned studies all confirmed the square root law (SRL), where the covert rate is $O(1/\sqrt{n})$ and approaches to $0$ as the number of channel uses $n$ goes to infinite. In order to achieve a positive covert rate, some existing studies tried to exploit channel uncertainties to improve the covert communication performance (e.g., \cite{Lee1,He1,Goeckel,Shahzad1, Wang2}). Since the statistical features and power of artificial noise (AN) are easy to control, jamming schemes are easier to implement in practice relative to the schemes exploiting channel uncertainties (e.g., \cite{SSobers2, AShahzad2, Shahzad3, Soltani,ZLiu}).  Furthermore, AN-aided schemes with poisson field random interferers were investigated to enhance the covert communication performance by considering location uncertainties\cite{TZheng}.  Most recently, covert communications with  multi-antenna interferers were examined in \cite{XChen, OShmuel}. Finally, the authors of  \cite{XChen} proposed to use multiple antennas with a full-duplex jammer to achieve covert communications against a warden Willie with uncertain locations. Meanwhile,
Shmuel \emph{et al.} in \cite{OShmuel} analyzed the effect of multiple antennas at the jammer on Alice's transmit power and consequently on the covert transmission rate.

Most of the works mentioned above were based on the assumption that the friendly jammer's location is fixed or quasi-static. This may cause some issues in practical scenarios. For instance, the static jammer cannot adaptively adjust its location according to the communication environment and thus may fail to provide robust opportunities and shields for secure or covert communications. Therefore, it has become increasingly urgent and necessary to use a mobile jammer for improving the security and privacy (i.e., covertness) of ground-based wireless communications in the presence of eavesdroppers or wardens. Actually, an unmanned aerial vehicles (UAV) is very suitable for serving as such a mobile jammer, due to its flexibility, mobility and fast deployment in practical applications. Against this background, UAVs have been extensively used for enhancing physical layer security against eavesdropping attacks in the literature. For example, the authors of \cite{Zhang1} jointly optimized the UAV's trajectory and transmit power in the presence of an eavesdropper to enhance the system secrecy performance. In addition, \cite{Gao} investigated the downlink UAV secure communication, where the UAV is prohibited from crossing the no-fly zones. In this work, the authors designed the UAV's trajectory and transmit power over a given flight period to maximize the average secrecy rate. Furthermore, the authors of \cite{Wang3} and \cite{Wang4} jointly optimized the location and transmit power of a UAV to enhance the security of UAV-enabled relaying networks. Meanwhile, the strategy of using a UAV as a friendly jammer to transmit AN, aiming to prevent an eavesdropper from eavesdropping on the confidential information, was considered in the literature (e.g.,  \cite{Cai,Li[],Zhou4,Lee2,YZhou,YCai}). These works demonstrated that the strategy of using UAVs as jammers can efficiently enhance physical layer security.

Motivated by the performance gain achieved by using UAV in the context of physical layer security, UAV-aided covert communications have also been attracting an increasing amount of research interests. The work \cite{Zhou1} initially considered UAV-aided covert communications, where the UAV acts as the transmitter for sending information to the legitimate receiver Bob covertly,  avoiding to be detected by the on-ground warden Willie. In this work, the authors considered the location randomness at both Bob and Willie for analyzing the covert communication performance. Meanwhile, the authors of \cite{HWang} considered a UAV as the warden Willie and a multi-hop relaying strategy was optimized to maximize the throughput of a ground network subject to the covertness constraint. In addition, a full-duplex UAV was employed to collect critical information from ground users in \cite{Zhou2}, where the UAV also generates AN with random transmit power to aid hiding the scheduled user's transmission from the unscheduled users. Furthermore, the covertness of a UAV based wireless transmission in the scenario where the UAV conducts surveillance (e.g., imaging) over a geographic area was considered in \cite{SYan}. Most recently, the work \cite{XJiang} considered a UAV-assisted multi-user covert data dissemination and jamming-assisted ground-air covert communication with multiple wardens. Based on \cite{XJiang}, covert communication
was further investigated in \cite{XXChen} with a multiple antennas at both the transmitter and the jammer
to fight against randomly distributed wardens in a UAV-aided network, where  a multi-antenna jammer is to maximize the transmission rate between a ground transmitter and a UAV receiver against several randomly distributed wardens.

In the aforementioned works on UAV-aided covert communications, the design of the UAV's trajectory or location is critical, since in general the UAV's trajectory or location affects both the communication performance from the transmitter Alice to the receiver Bob and the detection performance at the warden Willie. Therefore, the optimal design of the UAV's trajectory or location should balance between the communication and detection performance. Due to the non-convexities of the resultant optimization problems for designing the UAV's trajectory in covert communications, some traditional methods, e.g., block coordinate descent (BCD) in \cite{Li[]} and penalty successive convex approximation (P-SCA) in \cite{Zhou2}, were widely used to tackle the solutions to such non-convex optimization problems. However, there are generally two issues associated with these methods. The first one is that these methods widely relay on multiple approximations based on Taylor expansions, which limits the accuracy and optimality of the obtained solutions. The second issue is that the solutions achieved using these methods highly depend on the UAV's trajectory initialization, which is hard to be optimized. Against this background, in this work we considered a UAV-assisted covert communication system, where the UAV is employed as a dedicated mobile jammer to deliberately create interference at Willie for aid the covert communication from Alice to Bob. To resolve the aforementioned two issues, we develop, for the first time, a geometric method (GM) method to tackle the joint optimal design of the UAV's trajectory and Alice's transmit power in this work, which does not require the approximations based Taylor expansions or the UAV's trajectory initialization.
The main contributions of this work are summarized as below.

\begin{itemize}

\item As an early attempt to design UAV-assisted covert communications, we employ an UAV as a mobile jammer to broadcast AN signals with random transmit power for enhancing the covert communication performance. The UAV not only creates interference to Willie, but also to Bob. To optimize the system design, we develop a framework of jointly optimizing the UAV's trajectory and Alice's transmit power, aiming to extend the applications of covert communications from static application scenarios to dynamic ones.

\item For the first time, we develop a GM method to solve the joint optimization problem of the UAV's trajectory and Alice's transmit power. The development is based on the property of the formulated optimization problem, i.e., the equality of covertness constraint is always satisfied in the solution, which enables us to decouple the joint design into designing the UAV's trajectory and Alice's transmit power separately. Exploiting some properties associated with the Apollonius of Sphere, we prove that the UAV's optimal trajectory can be identified by solving an equation set efficiently, based on which Alice's optimal transmit power can be obtained in a closed-form expression. We note that our developed GM method does not require approximations based on Taylor expansions or the initialization of the UAV's trajectory, which totally avoids the widely existing issues in solving such optimization problems.

\item In order to demonstrate the benefits of our developed GM method, we also detail a conventional iterative (CI) method as a  benchmark scheme to tackle the joint design optimization problem, where the UAV's trajectory initialization is determined in two cases. Comparing our GM method with the CI benchmark method, we demonstrate that our developed GM method can achieve a significant covert performance gain in terms of achieving a higher average covert rate in the considered UAV-aided covert communication system. In addition, our examination shows that, as long as the UAV's flight time is sufficient, the UAV's hovering location (which is a critical location in the UAV's optimal trajectory) is always on the ray from Bob to Willie. Furthermore, our complexity analysis indicates that our developed GM method is of a lower complexity than the CI method.

\end{itemize}

The rest of this paper is organized as follows. In Section II, we detail the system model and the optimization problem. Then, we develop the GM method in details to solve the optimization problem in Section III. Subsequently, to demonstrate the benefit of our developed GM method, we detail the benchmark CI method in Section IV.
We present numerical results in Section V. Finally, conclusions are drawn in Section VI.


\begin{table*}[htb]
\centering  \caption{COMMONLY USED NOTATION}
\begin{tabular}{|p{0.15\textwidth}|p{0.55\textwidth}|l|}  
\hline
$\mathbb{E}_{x}[\cdot]$ &The
statistical expectation of $x$ \\

$\|\cdot\| $ & Euclidean norm\\
$|\cdot|$ & Absolute value\\
$T$ & UAV's flight period\\
$\mathbb{P}(\cdot)$ & The probability of
an event\\
$\beta_{0}$ & The channel power gain at the reference distance of $1$ m\\
${P_a}[n]$ & The transmit power of
Alice in the $n$-th time slot\\
$\mathbf{q}_{u}[n]$& The continuous UAV's trajectory  at $n$-th time slot\\
$P_u[n]$ &  The jamming power of the UAV at $n$-th time slot\\
$\hat{P}_{u}[n]$ &  The maximum transmit power of the UAV at $n$-th time slot\\
$\alpha$ & The path loss exponent\\
$d_{i,j}$, & The corresponding distance between $i$ to $j$\\
$\mathcal{CN}(0,\sigma^{2})$, & The circularly symmetric Gaussian distribution with zero mean and variance $\sigma^{2}$.\\
\hline
\end{tabular}
\label{tab:2}
\end{table*}

\section{System Model and Problem Formulation}
\subsection{System Model}
\begin{figure}[!t]
\centering
\includegraphics[width=0.6\textwidth]{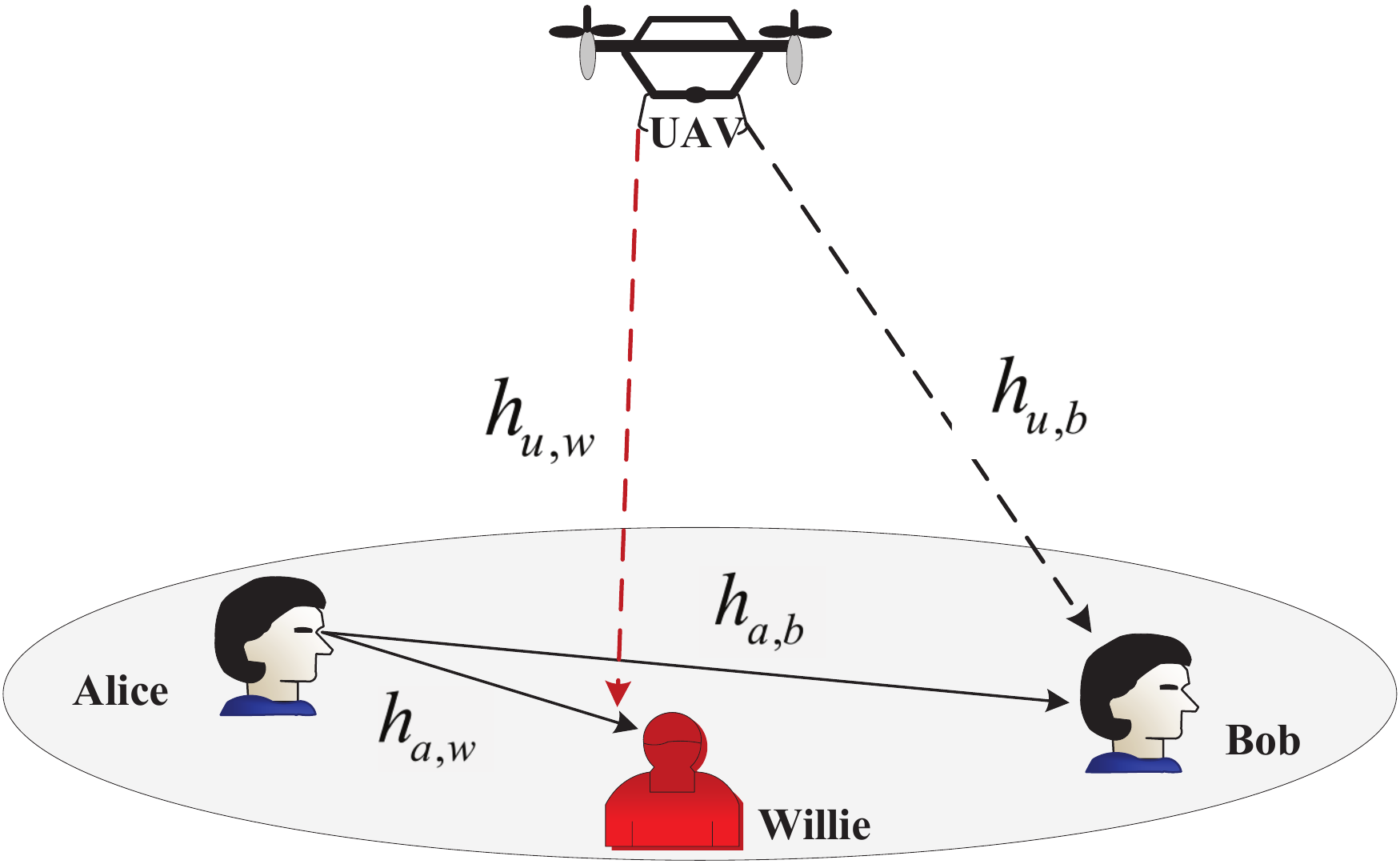}
\caption{A scenario of UAV-assisted covert wireless communications.}
\vspace{-0.5cm}
\label{fig:2}
\end{figure}

In this work, we consider a UAV-enabled covert communication system, where Alice, Bob and Willie are terrestrial users, while a UAV acts as a mobile jammer to aid the covert communication from Alice to Bob against the detection of Willie. Each of the UAV and all terrestrial users is equipped with a single antenna. In addition, the locations of Alice, Bob and Willie are fixed to $\mathbf{q}_{a}\triangleq[x_{a},y_{a}]^{T}$, $\mathbf{q}_{b}\triangleq[x_{b},y_{b}]^{T}$ and $\mathbf{q}_{w}\triangleq[x_{w},y_{w}]^{T}$, respectively, while the UAV flies at a fixed altitude of $H$. To facilitate
the UAV trajectory design, we equally divide the flight period $T$ into $N$ time slots with the duration $\sigma_{t}=\frac{T}{N}$, and $N$ is large enough such that the distances between the UAV and the terrestrial  users
are approximately constants within each time slot. Thus, the continuous UAV's trajectory  can be replaced by $N$ discrete positions $\mathbf{q}_{u}[n]\triangleq[x_{n},y_{n}]^{T}$, and $n\in \mathcal{N}\triangleq\{0, 1, 2, \ldots, N\}$ is the index of time slot.

We assume that the instantaneous channel state information (CSI) of all involved channels and the locations of terrestrial users are available to the UAV. The channels from Alice to Bob and from Alice to Willie are subject to Rayleigh fading. Thus, the corresponding channel power gains can be written as  $|h_{a,b}[n]|^{2}=\beta_{0}d_{a,b}^{-\alpha}\zeta_{a,b}[n]$  and  $|h_{a,w}[n]|^{2}=\beta_{0}d_{a,w}^{-\alpha}\zeta_{a,w}[n]$, respectively, where $\beta_{0}$ is the channel power gain at the reference distance of $1$ m, $d_{a,b}$ and $d_{a,w}$ are the corresponding distances between two transceivers, $\alpha$ is the path loss exponent, $\zeta_{a,b}[n]$ and $\zeta_{a,w}[n]$ are exponentially distributed random variables with the parameter $\lambda$ = 1. Moreover, for the reader's convenience,  Table I summarized all notations used in the paper.

Following \cite{Wu[]}, we note that the UAV-ground channels are mainly dominated by line-of-sight (LoS) components. Thus, the UAV can practically obtain the CSI of the channels to the ground users (i.e., Alice, Bob, and Willie) if the location information of these ground users is known.  In this work, we assume that the
UAV knows the  exact locations of the ground users, e.g., through a surveillance or based on the history communication information to these users \cite{Lee2}. In addition, this assumption has been also widely used in existing works on UAV assisted covert communication (e.g., \cite{Zhou1},\cite{Zhou2} and \cite{Yan[]}).
 Accordingly, for the $n$-th time slot, the channel power gains from the UAV to Bob and Willie can be expressed as
\begin{equation}\label{mod:1}
\begin{aligned}
|h_{u,b}[n]|^{2}=\frac{\beta_{0}}{\|\mathbf{q}_{u}[n]-\mathbf{q}_{b}\|^{2}+H^{2}}, \\
\end{aligned}
\end{equation}
and
\begin{equation}\label{mod:2}
\begin{aligned}
|h_{u,w}[n]|^{2}=\frac{\beta_{0}}{\|\mathbf{q}_{u}[n]-\mathbf{q}_{w}\|^{2}+H^{2}}, \\
\end{aligned}
\end{equation}
respectively.

We denote
 $x_{a}^{(i)}[n]\sim\mathcal{CN}(0,1)$ and ${P_a}[n]$ as the transmitted symbol and transmit power of Alice for the $i$-th channel use  ($i=1,2...,m$) in the $n$-th time slot, respectively, where $m$ is the total number of channel uses in each time slot. We also denote $v_{b}^{(i)}[n]\sim \mathcal{CN}(0,\sigma_{b}^{2})$ as the corresponding AWGN at Bob. Then, the received signals at Bob can be written as

\begin{equation}\label{mod:3}
y_{b}^{(i)}[n]=\sqrt{P_{a}[n]}h_{a,b}[n]x_{a}^{(i)}[n]+\Phi[n]+v_{b}^{(i)}[n],
\end{equation} where  $\Phi[n]=\sqrt{P_{u}[n]}h_{u,b}[n]x_{u}^{(i)}[n]$, $P_u[n]$ denotes the jamming power of the UAV, and $x_{u}^{(i)}[n]$ denotes the transmitted AN symbol of the UAV satisfying $\mathbb{E}[x_{u}^{(i)}x_{u}^{(i)\dagger}] =1$.
Similar to \cite{SSobers2}, we assume that ${P}_{u}[n]$ is an independent identically distributed (i.i.d.) uniform random variable with the value range $[0, \hat{P}_{u}[n]]$,  where $\hat{P}_{u}[n]$  is the maximum transmit power of the UAV at each time slot.  Specifically, the probability density function (PDF) of $P_{u}[n]$ is given by
\begin{equation}\label{mod:4}
f_{P_{u}[n]}(x)=
\begin{cases}
\frac{1}{\hat{P}_{u}[n]},&\text{if}~0\leq x\leq \hat{P}_{u}[n], \\
0, &\text{otherwise}.
\end{cases}
\end{equation}
Based on \eqref{mod:3}, the channel capacity from Alice to Bob is given by
\begin{equation}\label{mod:5}
\begin{aligned}
C=\log_{2}\left(1+\frac{P_{a}[n]\beta_{0}d_{a,b}^{-\alpha}\zeta_{a,b}[n]}{P_{u}[n]|h_{u,b}[n]|^{2}+\sigma_{b}^{2}}\right).
\end{aligned}
\end{equation}
We assume that Bob knows the maximum transmit power of the UAV at each time slot, but he does not know $P_{u}[n]$. Thus, a communication outage occurs when $C$ cannot support the transmission rate. The outage probability can be expressed as
\begin{eqnarray}\label{mod:6}
\begin{aligned}
\mathbb{P}_{out}=\mathbb{P}\{C\leq R_{b}[n]\}=1-\exp\left(-\nu[n]\right),
\end{aligned}
\end{eqnarray}
where
\begin{eqnarray}\label{mod:8}
\begin{aligned}
\nu[n]=\frac{(2^{R_{b}[n]}-1)(P_{u}[n]|h_{u,b}[n]|^{2}+\sigma_{b}^{2})}{P_{a}[n]\beta_{0}d_{a,b}^{-\alpha}}
\end{aligned}
\end{eqnarray}
and $R_{b}[n]$ is the transmission rate from Alice to Bob.

From \eqref{mod:6}, we can see that the outage probability monotonically increases
with $P_{u}[n]$. Then, if the outage probability is guaranteed to be always lower than or equal to a threshold $\rho_{b}$ for all possible $P_{u}[n]$, $R_{b}[n]$ is given by
\begin{eqnarray}\label{mod:11}
\begin{aligned}
R_{b}[n]\leq\log_{2}\left(1-\frac{\ln(1-\rho_{b})P_{a}[n]\beta_{0}d_{a,b}^{-\alpha}}{\hat{P}_{u}[n]|h_{u,b}[n]|^{2}+\sigma_{b}^{2}}\right).
\end{aligned}
\end{eqnarray}

\subsection{Detection Performance at Willie}
The received signal at Willie can be written as

\begin{equation}\label{mod:12}
y_{w}^{(i)}[n]=\left\{\aligned &\Delta[n]+v_{w}^{(i)}[n], &H_{0}, \\
&\sqrt{P_{a}[n]}h_{a,w}[n]x_{a}^{(i)}[n]+\Delta[n]+v_{w}^{(i)}[n], &H_{1},
\endaligned\right.
\end{equation}
where  $y_{w}^{(i)}[n]$ denotes the received signal at Willie for
the $i$-th channel use in the $n$-th time slot,
$\Delta[n]=\sqrt{P_{u}[n]}h_{u,w}[n]x_{u}^{(i)}[n]$, $H_{0}$ is the null hypothesis in which Alice did not
transmit, $ H_{1}$ is the alternative hypothesis where Alice did
transmit to Bob, and $v_{w}^{(i)}\sim\mathcal{CN}(0,\sigma_{w}^{2})$ is the AWGN at Willie.

In this work, following \cite{SSobers2} we assume that Willie uses a radiometer detector at each time slot to detect the transmission from Alice to Bob.  Willie decides on the state
$H_{1}$  if ${{P}_{w}}[n]\geq \vartheta$. Otherwise, Willie decides on the state $H_{0}$. Therefore, we have the decision rule as
\begin{equation}\label{mod:13}
\begin{aligned}
{{P}_{w}}[n]\triangleq \frac{1}{m}\ \sum\limits_{i=1}^{m}{{{\left| {{y}_{w}}^{(i)}[n] \right|}^{2}}}\overset{H_{1}}{\underset{H_0}{\gtrless}} \vartheta,
\end{aligned}
\end{equation}
where $P_{w}[n]$ is the average received power at each time slot and $\vartheta$ is the threshold in the detector.

We denote $\mathbb{P}_{FA}[n]$ and $\mathbb{P}_{MD}[n]$ as the false alarm rate and the miss
detection rate, respectively. Then, as per  \eqref{mod:12} we have
\begin{align}
\mathbb{P}_{FA}[n]&=\mathbb{P}\left (P_{w}[n]>\vartheta |H_{0}\right)=\mathbb{P}\left((\sigma_{w}^{2}+\gamma[n])\frac{\mathcal{X}_{2m}^{2}}{m}>\vartheta|H_{0}\right),\label{mod:14}\\
\mathbb{P}_{MD}[n]&=\mathbb{P}\left(P_{w}[n]<\vartheta|H_{1}\right)=\mathbb{P}\left((\sigma_{w}^{2}+\gamma[n])\frac{\mathcal{X}_{2m}^{2}}{m}<\vartheta|H_{1}\right),\label{mod:15}
\end{align}
where $\mathcal{X}_{2m}^{2}$ represents the chi-squared random variable with $2m$ degrees of freedom and
\begin{eqnarray}\label{mod:15A}
\gamma[n]=
\begin{cases}
{P}_{u}[n]|h_{u,w}[n]|^{2},&H_{0} , \\
\beta_{0}P_{a}[n]d_{a,w}^{-\alpha}\zeta_{a,w}[n]+{P}_{u}[n]|h_{u,w}[n]|^{2}, &H_{1}.\\
\end{cases}
\end{eqnarray}
Meanwhile, based on the strong law of large numbers, we have
\begin{eqnarray}\label{mod:15AA}
\begin{aligned}
\Pr \left\{\lim_{m\rightarrow\infty} \frac{\mathcal{X}_{2m}^{2}}{m}=1\right\}=1.
\end{aligned}
\end{eqnarray}
Therefore, the expression for $\mathbb{P}_{FA}[n]$ and $\mathbb{P}_{MD}[n]$ are given in the following lemma.
\begin{lemma}\label{Lemma:1}
According to \eqref{mod:14} and \eqref{mod:15}, the false
alarm and miss detection rates in the $n$-th time slot at Willie are given by
\begin{eqnarray}\label{mod:16}
\mathbb{P}_{FA}[n]=
\begin{cases}
1,&\vartheta-\sigma_{w}^{2}\leq0 , \\
1-\frac{\vartheta-\sigma_{w}^{2}}{\Gamma[n]}, &0<\vartheta-\sigma_{w}^{2}\leq\Gamma[n],\\
0,&\vartheta-\sigma_{w}^{2}>\Gamma[n],
\end{cases}
\end{eqnarray}
and
\begin{eqnarray}\label{mod:17}
\mathbb{P}_{MD}[n]=
\begin{cases}
 0,&\vartheta-\sigma_{w}^{2}\leq0 , \\
\Lambda[n], &0<\vartheta-\sigma_{w}^{2}\leq\Gamma,\\
\Omega[n], &\vartheta-\sigma_{w}^{2}>\Gamma,
\end{cases}
\end{eqnarray}
where $\Gamma[n]=\hat{P}_{u}[n]|h_{u,w}[n]|^{2}$, $\phi = \frac{{{\beta _0}}}{{d_{a,w}^\alpha }}$,
\begin{align}
\Lambda[n]  = \frac{{\phi {P_a}[n]}}{\Gamma[n] }\exp \left( { - \frac{{(\vartheta  - \sigma _w^2)}}{{\phi {P_a}[n]}} - 1} \right) + \frac{{\vartheta  - \sigma _w^2}}{\Gamma[n] },\notag
\end{align}
and
\begin{equation}
\Omega[n]  = 1 - \frac{{\phi {P_a}[n]}}{\Gamma[n] }\left( {\exp \left( {\frac{{\sigma _w^2 + \Gamma[n]  - \vartheta }}{{\phi {P_a}[n]}}} \right) - \exp \left( {\frac{{\sigma _w^2 - \vartheta }}{{\phi {P_a}[n]}}} \right)} \right).\notag
\end{equation}
\end{lemma}
\begin{IEEEproof}
The detailed proof is similar to that of Lemma~1 in \cite{Zhou2}.
\end{IEEEproof}

Based on the above lemma, the minimum value of the detection error rate, which is defined as $\xi=P_{FA}+P_{MD}$,  and corresponding optimal detection threshold that minimizes $\xi$ are given in the following lemma.

\begin{lemma}\label{Lemma:2}
For the radiometer detector given in \eqref{mod:13}, the optimal detection threshold at  Willie in the $n$-th time slot is $\vartheta^{\ast}=\Gamma[n]+\sigma_{w}^{2}$, and the minimum detection error rate is given by
\begin{equation}\label{mod:20}
{\xi ^*}[n] = 1 + \left( {\exp \left( { - \frac{\Gamma[n] }{{\phi {P_a}[n]}}} \right) - 1} \right)\frac{{\phi {P_a}[n]}}{\Gamma[n] } .
\end{equation}
\end{lemma}
\begin{IEEEproof}
The proof is similar to that of Theorem 1  in \cite{Zhou2}.
\end{IEEEproof}

\subsection{Optimal Design Problem Formulation}

In this work, we aim to maximize the transmission rate from Alice to Bob subject to the covertness and reliability constraints via a joint design of the UAV's trajectory and Alice's transmit power. Defining $\mathbf{Q}\triangleq\{\mathbf{q}[n],\forall n\}$ and $\mathbf{P}\triangleq\{P_{a}[n],\forall n\}$, the optimization problem for the optimal design is formulated as

\begin{subequations}\label{mod:18}
\begin{align}
{\bf (P1):~~}&\underset{\mathbf{Q},\mathbf{P}}{\text{maximize}}~~\mathcal{L}(\mathbf{Q},\mathbf{P})= \frac{1}{N}\sum_{n=1}^{N}R_{b}[n] \label{18a}\\
\text{s.t. ~}
& \xi^{\ast}[n]\geq1-\epsilon\label{18b},\\
&\|\mathbf{q}_{u}[n]-\mathbf{q}_{u}[n-1]\|\leq V_{\max}\sigma_{t}, \forall n,\label{18c}\\
&\mathbf{q}_{u}[N]=\mathbf{q}_{u,F},\label{18d}\\
&\mathbf{q}_{u}[0]=\mathbf{q}_{u,0},\label{18e}
\end{align}
\end{subequations}
 where $V_{\max}$ is the maximum speed of the UAV, $\mathbf{q}_{u,0}=[x_{u,0}, y_{u,0}]$ and $\mathbf{q}_{u,F}=[x_{u, F}, y_{u, F}]$ are the initial and final locations of the UAV, respectively.

We note that constraint \eqref{18b} ensures the covertness of the transmission from Alice to Bob and constraints \eqref{18c}, \eqref{18d}, and \eqref{18e} are the UAV's mobility constraints. We also note that \eqref{mod:18} is non-convex due to the non-convex structures of both the objective function and constraints. This leads to that the above optimization problem is difficult to be solved directly. Therefore, we first develop the GM to solve it in the following section and then we also present the CI method as a benchmark to solve it based on BCD and concave-convex procedure (CCCP) in Section IV.

\section{Geometric method to Solve the Optimization Problem}
In this section, we aim to to design Alice's transmit power and UAV's  trajectory to ensure that Alice can transmit signal to Bob  reliably and covertly. To this end, we first
equivalently decompose the optimization problem \eqref{mod:18} and then develop a GM algorithm to solve the resultant problems, which can strike a balance between the covert communication performance and computational complexity. The key idea of the GM algorithm is to
first analyze the geometrical features of the UAV's optimal trajectory and then determine Alice's optimal transmit power as per the obtained optimal trajectory.

\subsection{Problem Decomposition}

To apply the GM algorithm, we first decompose the optimization problem into two subproblems in this subsection. To proceed,
 following \eqref{mod:11} we first approximately equivalently rewrite the optimization problem \eqref{mod:18} as
 \begin{subequations}\label{mod:s1}
\begin{align}
&\underset{\mathbf{Q},\mathbf{P}}{\text{maximize}}~~\frac{1}{N}\sum_{n=1}^{N}\left(\frac{\left(d_{u,b}[n]\right)^{2}}{\left(d_{u,w}[n]\right)^{2}}\times \kappa[n] P_{a}[n]\tau[n]\right) \label{19a}\\
\text{s.t. ~}
&\eqref{18b}, \eqref{18c}, \eqref{18d}, \eqref{18e},
\end{align}
\end{subequations}
where $d_{u,b}[n]=\sqrt{\|\mathbf{q}_{u}[n]-\mathbf{q}_{b}\|^{2}+H^{2}}$, $d_{u,w}[n]=\sqrt{\|\mathbf{q}_{u}[n]-\mathbf{q}_{w}\|^{2}+H^{2}}$, $\tau[n]=\frac{\left(d_{u,w}[n]\right)^{2}}{\hat{P}_{u}[n]d_{a,w}^{\alpha}}$ and $\kappa[n]=-\frac{\ln(1-\rho_{b})d_{a,b}^{-\alpha}d_{a,w}^{\alpha}}{\hat{P}_{u}[n]}$. In addition, as per \eqref{mod:20}, the constraint \eqref{18b} can be rewritten as
\begin{align}\label{s20}
{\rm{G}}\left( {{P_a}\left[ n \right],{{\bf{q}}_u}\left[ n \right]} \right)= P_{a}[n]\tau[n]\left(1-e^{-\frac{1}{P_{a}[n]\tau[n]}}\right)\leq \epsilon.
\end{align}
In order to provide a theoretical support for the decomposition of  \eqref{mod:s1}, we first present the following lemma on the monotony of ${\rm G}\left( {{P_a}\left[ n \right],{{\bf{q}}_u}\left[ n \right]} \right)$ with respect to ${P_a}[n]$.

\begin{lemma}\label{lemma:3}
For a given ${{\bf{q}}_u}\left[ n \right]$, ${\rm G}\left( {{P_a}\left[ n \right],{{\bf{q}}_u}\left[ n \right]} \right)$ is a monotonically increasing function of $P_{a}[n]$.
\end{lemma}

\begin{IEEEproof}
The detailed proof is provided in  Appendix \ref{app:lemma:3}.
\end{IEEEproof}

Following Lemma~\ref{lemma:3} and noting ${\rm G}\left( {{P_a}\left[ n \right],{{\bf{q}}_u}\left[ n \right]} \right)=0$ for $P_a\left[ n \right]=0$, we conclude that there always exists a feasible $\mathbf{P}$ satisfying the constraint \eqref{18b} for any feasible trajectory $\bf Q$ satisfying \eqref{18c}, \eqref{18d} and \eqref{18e}. This means that as long as the optimal trajectory $\mathbf{Q}$ exists and is determined, we can always identify a unique optimal value for Alice's transmit power $\mathbf{P}$ corresponding to the optimal trajectory.

As per Lemma~\ref{lemma:3} and noting that the objective function \eqref{19a} also monotonically increases with $P_{a}[n]$, we can conclude that the equality in the constraint \eqref{18b} must hold in the solution to the optimization problem \eqref{mod:s1}. Then, as per \eqref{19a} we note that the UAV's trajectory $\mathbf{Q}$ only affects the first term in \eqref{19a}.

Inspired by the above conclusion, we decompose the optimization problem \eqref{mod:s1} into the following two subproblems:
\begin{subequations}\label{mod:ss2}
\begin{align}
{\bf (P1.1):~}&\underset{\mathbf{Q}}{\text{maximize}}~~ \frac{1}{N}\sum_{n=1}^{N} \left(\frac{\left(d_{u,b}[n]\right)^{2}}{\left(d_{u,w}[n]\right)^{2}}\right)\label{21a}\\
\text{s.t. ~}
&\|\mathbf{q}_{u}[n]-\mathbf{q}_{u}[n-1]\|\leq V_{\max}\sigma_{t}, n =1,...,N,\label{21b}\\
&\mathbf{q}_{u}[N]=\mathbf{q}_{u,F},\label{21c} \\
&\mathbf{q}_{u}[0]=\mathbf{q}_{u,0},\label{21d}
\end{align}
\end{subequations}
and
\begin{subequations}\label{mod:ss3}
\begin{align}
{\bf (P1.2):~}&\underset{\mathbf{P}}{\text{maximize}} ~~\frac{1}{N}\sum_{n=1}^{N} \left(\eta P_{a}[n]\tau[n] \right) \\
\text{s.t. ~}
&P_{a}[n]\tau[n]\left(1-e^{-\frac{1}{P_{a}[n]\tau[n]}}\right)\leq \epsilon,\label{22b}
\end{align}
\end{subequations}
where $\eta=-\ln(1-\rho_{b})d_{a,b}^{-\alpha}$.

Based on the objective function \eqref{21a}, we note that the UAV prefers the positions that are close to Willie but far from Bob, in order to support the covert data transmission. Intuitively, this is due to that the UAV can avoid creating interference to Bob by keeping far away from him, while the UAV can create more interference to Willie for satisfying the covertness constraint if it keeps close to Willie. From \eqref{22b}, Alice's optimal
transmit power $P_{a}[n]$ increases with the UAV's maximum transmit power ${\hat P}_{u}\left[ n \right]$.  This is due to the fact that the equality in the covertness constraint always holds in the solution (i.e., $P_{a}[n]\tau[n]\left(1-e^{-\frac{1}{P_{a}[n]\tau[n]}}\right)= \epsilon$), while $\tau[n]$ monotonically decreases with ${\hat P}_{u}\left[ n \right]$ and $P_{a}[n]\tau[n]\left(1-e^{-\frac{1}{P_{a}[n]\tau[n]}}\right)$ increases with $P_{a}[n]$ as per Lemma~\ref{lemma:3}.
In addition, we also note that Alice's optimal transmit power increases as the UAV moves closer to the Willie (i.e., $d_{uw}[n]$ decreases) as per the definition of $\tau[n]$.



Based on the above analysis, in the following subsection, we prove that there are unique optimal solutions to the subproblems {\bf (P1.1)} and {\bf (P1.2)}.
This motivates us to develop the GM algorithm to solve the optimization problems {\bf (P1.1)} and {\bf (P1.2)}, which can balance between the computational complexity and the achievable covert communication performance.


\subsection{Geometric Scheme for Solving (\ref{mod:ss2}): UAV's Trajectory Design}

\begin{figure}[!h]
\centering
\includegraphics[width=0.5\textwidth]{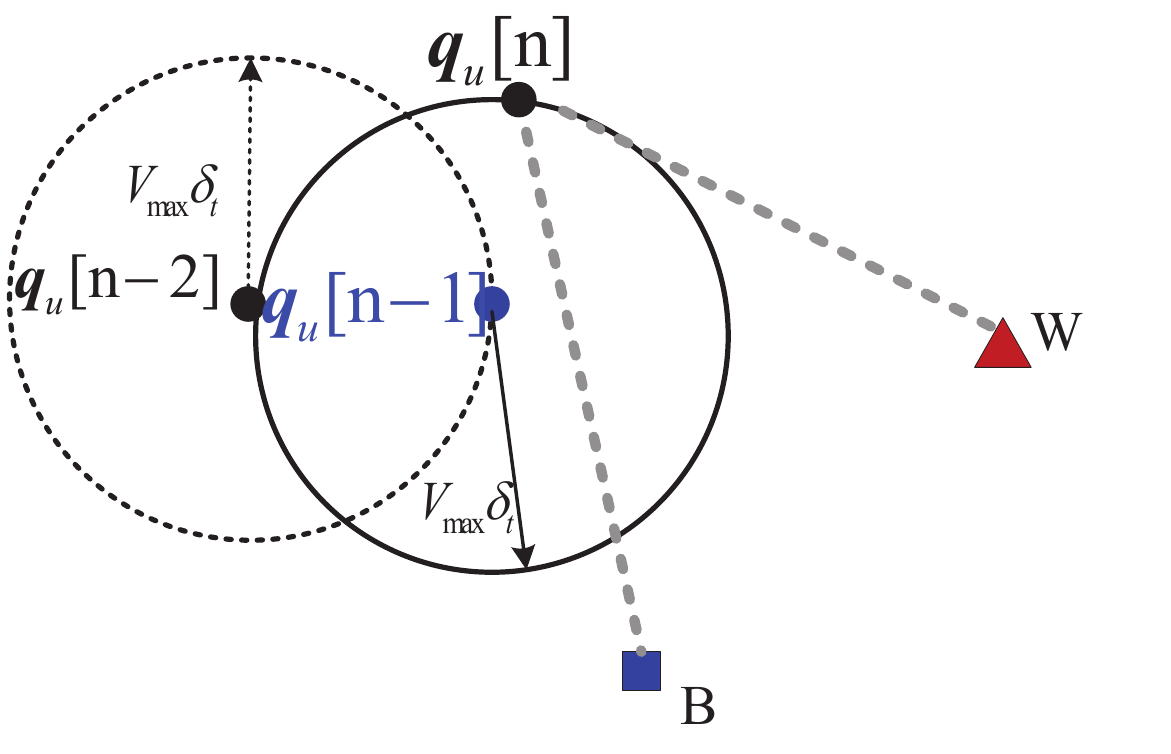}
\caption{The feasible region of the UAV's trajectory in each time slot.}
\label{fig:7}
\end{figure}

In this subsection, we develop a GM method for the UAV's trajectory design. Based on \eqref{21b}, we know that the feasible ${\bf q}_u[n]$, $0<n<N$, must be within the plane centered at ${\bf q}_u[n-1]$ with the radius of $V_{\max}\delta_t$, as shown in Fig.~\ref{fig:7}. In this subsection, we propose to design the UAV trajectory slot by slot instead of optimizing the trajectory across all the time slots.

We first note that the constraint  \eqref{21c} of the optimization problem \eqref{mod:ss2} requires that the UAV returns to ${\bf {q}}_{u,F}$ at the end of the last time slot. Therefore, the remaining $(N-n)$ time slots should be enough for the UAV directly flying from ${\widehat {\bf{q}}_u}[n]$ to ${\bf {q}}_{u,F}$, where ${\widehat {\bf{q}}_u}[n]$ is the turning point for the UAV to return to the destination. If this condition is not satisfied, ${\bf q}_u[n]$ should be on the straight line between ${{\bf{q}}_u}[n-1]$ and ${\bf {q}}_{u,F}$. As a result, ${{\bf{q}}_u}[n]^\ast$ is given by
\begin{equation}\label{mod:v1}
{{\bf{q}}_u}[n]^\ast=
\begin{cases}
{\widehat {\bf{q}}_u}[n], &N-n\geq N_{1},\\
{\mathbf{q}_u[n-1]} + (n-N){V_{\max }}{\delta _t}\theta_{1},&
N-n<N_{1},
\end{cases}
\end{equation}
where $N_{1}=\frac{||\mathbf{q}_{u}[n]-\mathbf{q}_{u,F}||}{(V_{\max}\delta_{t})}$ and $\theta_{1}=\frac{\mathbf{q}_{u}[n-1]-\mathbf{q}_{u,F}}{||\mathbf{q}_{u}[n-1]-\mathbf{q}_{u,F}||}$.

As such, for the first $n$ time slots, $0<n<N$, the UAV's trajectory design problem can be formulated as
\begin{subequations}\label{new}
\begin{align}
 &\underset{\mathbf{q}_{u}[n]}{\text{maximize}}~~ \frac{{{d_{u,b}}[n]}}{{{d_{u,w}}[n]}} \buildrel \Delta \over = \sqrt {\frac{{{{\left\| {{{\bf{q}}_u}[n] - {{\bf{q}}_b}} \right\|}^2} + {H^2}}}{{{{\left\| {{{\bf{q}}_u}[n] - {{\bf{q}}_w}} \right\|}^2} + {H^2}}}},\label{24a}\\
&\text{s.t. ~}
\|\mathbf{q}_{u}[n]-\mathbf{q}_{u}[n-1]\|\leq V_{\max}\sigma_{t}.\label{24b}
\end{align}
\end{subequations}

Following  \eqref{24a} and  \eqref{24b}, we note that the essence of the optimization problem \eqref{new} is to find a point $\mathbf{q}_{u}[n]$ in the feasible region determined by the constraint  \eqref{24b}, to maximize the ratio from the distance between $\mathbf{q}_{u}[n]$ and Bob to the distance between $\mathbf{q}_{u}[n]$ and Willie. We present the following theorem to facilitate solving the optimization problem (\ref{new}).


\begin{theorem}\label{theorem:4}
Any point $\overline {\bf{q}} [n] = {\left[ {{x_n},{y_n},{z_n}} \right]^T}$ satisfying $\frac{{\left\| {\overline {\bf{q}} [n] - {{\bf{q}}_b}} \right\|}}{{\left\| {\overline {\bf{q}} [n] - {{\bf{q}}_w}} \right\|}} = k$ must be on the spherical surface of Apollonius of Sphere when $k \neq 1$ and on a line when $k = 1$, i.e., we have
\begin{equation}\notag
\left\{ {\begin{array}{*{20}{l}}
{{{\left( {{x_n} + x'} \right)}^2} + {{\left( {{y_n} + y'} \right)}^2} + {z_n}^2 = R{{\left( k \right)}^2},}&{k \ne 1,}\\
{{x_n}\left( {{x_w} - {x_b}} \right) + {y_n}\left( {{y_w} - {y_b}} \right) = 0,}&{k = 1,}
\end{array}} \right.
\end{equation}
where $x'=\frac{k^{2}x_{w}-x_{b}}{1-k^{2}}$, $y'=\frac{k^{2}y_{w}-x_{b}}{1-k^{2}}$, $R\left( k \right) = \frac{{k{{\left\| {{{\bf{q}}_b} - {{\bf{q}}_w}} \right\|}}}}{{{k^2} - 1}}$, and $k>0$.
\end{theorem}

\begin{IEEEproof}
The detailed proof is provided in Appendix \ref{app:theorem:4}.
\end{IEEEproof}

Based on Theorem~\ref{theorem:4}, the optimization problem \eqref{new} can be simplified to maximize $k$ subject to the constraints.
In addition, we find that $\mathop {\lim }\limits_{k \to 1} R\left( k \right) = \infty$, $\mathop {\lim }\limits_{k \to 0} R\left( k \right) = 0$, and $\mathop {\lim }\limits_{k \to \infty } R\left( k \right) = 0$. Therefore, there must exist a tangent point between the feasible region of ${\bf{q}}_u[n]$ and the Apollonius of Sphere. As a result, the optimal $k$ is presented in the following theorem.

\begin{theorem}\label{theorem:5}
The optimal $k$ that determines the solution to the optimization problem \eqref{new}, is given by
\begin{equation}\label{mod:v2}
{k^\ast} = \left\{ {\begin{array}{*{20}{l}}
{\sqrt {\frac{{{k_1} - \sqrt {k_1^2 - 4{k_0}{k_2}} }}{{2{k_0}}}} ,}&{{d_1} > {V_{\max }}{\sigma _t}\;{\rm{and}}\;{d_2} < {d_3}},\\
{1,}&{{d_1} = {V_{\max }}{\sigma _t}\;{\rm{and}}\;{d_2} < {d_3}},\\
{\sqrt {\frac{{{k_1} + \sqrt {k_1^2 - 4{k_0}{k_2}} }}{{2{k_0}}}} ,}&{{\rm{otherwise}},}
\end{array}} \right.
\end{equation}
where ${\bf{\tilde x}} = {{\bf{q}}_u}[n - 1] - {I^T}{(I{I^T})^{ - 1}}(I{{\bf{q}}_u}[n - 1] - \hat b)$, $I = {({{\bf{q}}_w} - {{\bf{q}}_b})^T}$, $\hat b = \frac{1}{2}{({{\bf{q}}_w} + {{\bf{q}}_b})^T}({{\bf{q}}_w} - {{\bf{q}}_b})$, $d_1={\|\mathbf{q}_{u}[n-1]-\mathbf{\tilde{x}}\|}$, $d_2=||\mathbf{q}_{u}[n-1]-\mathbf{q}_{b}||$, $d_3=||\mathbf{q}_{u}[n-1]-\mathbf{q}_{w}||$, $k_{0}=(U+S)^{2}+4(V_{\max}\delta_{t})^{2}H^{2}$, ${k_1} = {\left( {2SV - 2{U^2} - 2US + 2UV - 4{V_{\max }}{\delta _t}} \right)^2}\left( {{{\left\| {{{\bf{q}}_b} - {{\bf{q}}_w}} \right\|}^2} + {H^2}} \right)$, $U = {\left\| {{{\bf{q}}_u}[n - 1]} \right\|^2} - {({V_{\max }}{\delta _t})^2} + {H^2}$, $S=\|\mathbf{q}_{w}\|^{2}-\mathbf{q}_{w}^{T}\mathbf{q}_{u}[n-1]$, and $V=\|\mathbf{q}_{b}\|^{2}-\mathbf{q}_{b}^{T}\mathbf{q}_{u}[n-1]$.
\end{theorem}

\begin{IEEEproof}
The detailed proof is provided in Appendix \ref{app:theorem:5}.
\end{IEEEproof}

With $k^\ast$, the solution of ${{\bf{q}}_u}[n]^\ast$ to the optimization problem \eqref{new} can be obtained by solving
\begin{equation}\label{mod:v3}
\left\{ {\begin{array}{*{20}{l}}
{\left\| {{{\bf{q}}_u}[n]^\ast - {{\bf{q}}_u}\left[ {n - 1} \right]} \right\| = {V_{\max }}{\delta _t},}\\
{{{\left\| {{{\bf{q}}_u}[n]^\ast - {{\bf{q}}_o}} \right\|}^2} = R{{\left( k \right)}^2} - {H^2},}
\end{array}} \right.
\end{equation}
where ${{\bf{q}}_o} = {\left[ {x',y'} \right]^{\rm{T}}}$, $x'=\frac{k^{2}x_{w}-x_{b}}{1-k^{2}}$ and  $y'=\frac{k^{2}y_{w}-x_{b}}{1-k^{2}}$.

\subsection{Geometric Scheme for Solving \eqref{mod:ss3}: Alice's Transmit Power Design}
We note that, for a given $\mathbf{q}_{u}[n]$, $\forall n$, the optimization problem \eqref{mod:ss3} is a convex optimization problem, since the objective function
\begin{equation}\label{mod:v4}
\underset{\mathbf{P}}{\text{maximize}} ~~\frac{1}{N}\sum_{n=1}^{N} \left(\eta P_{a}[n]\tau[n] \right)
\end{equation}
monotonically increases with $P_{a}[n]$, which is a concave function of $P_{a}[n]$ and the covertness constraint
\begin{equation}\label{mod:v5}
P_{a}[n]\tau[n]\left(1-e^{-\frac{1}{P_{a}[n]\tau[n]}}\right)\leq \epsilon
\end{equation}
is convex with respect $P_{a}[n]$. Thus, the optimization problem
(\ref{mod:ss3}) can be efficiently solved by a convex optimization solver
such as CVX \cite{Boyd}.

%
%
%
%
%
%
%

Following the above discussions, the GM method can be summarized in {\bf Algorithm 1}.

\begin{algorithm}[htb]

\caption{GM method for Trajectory and Power Optimization}

\label{alg:Framwork}                  

\begin{algorithmic}[1]                

\REQUIRE ~~\\
\STATE \textbf{Initialize:}\\                         
\STATE~$n=1$, ${\bf{q}}_u[0]={\bf q}_{u,0}$;
\STATE \textbf{Repeat:}
\STATE~Solve problem \eqref{new} and obtain ${{\bf{q}}_u}[n]^\ast$ via (\ref{mod:v3});
\STATE~$n=n+1$;
\STATE \textbf{Until:}
 $n=N-1$;
 \STATE~${\bf{q}}_u[N]={\bf q}_{u,F}$;
 \STATE~Substitute $\mathbf{Q}^{\ast}$ into problem \eqref{mod:ss3}, then solve problem \eqref{mod:ss3} by CVX, and obtain the optimal solution is $\mathbf{P}^{\ast}$; \label{step:2}
\end{algorithmic}
\end{algorithm}


We next briefly analyze the complexity of Algorithm~1.
The complexity of GM for solving \eqref{mod:18} consists of the following two parts.

\begin{itemize}
\item [$\bullet$] The complexity of solving the optimization problem \eqref{mod:ss2}:
The optimization problem \eqref{mod:ss2} requires $N$ iterations to obtain ${\bf Q}^*$. Thus, its complexity is $\mathcal{O}(N)$.

\item [$\bullet$] The complexity of solving the optimization problem \eqref{mod:ss3}: We recall that the convex restrictions in \eqref{22b} only involve linear matrix inequality (LMI) constraints.  According to \cite{KYWang}, the essence of optimization problem \eqref{mod:ss3} is to identify the searching direction by solving $N$ linear equations with only one unknown variable in each equation. Thus, the complexity is dominated by the $N\times1$ coefficient matrix of the linear
system which is on the order of $\mathcal{O}(N^{4})$.
\end{itemize}

Considering the aforementioned two parts, the complexity
of the GM method for solving \eqref{mod:18} is on the order of $\mathcal{O}(N)+\mathcal{O}(N^{4})\triangleq\mathcal{O}(N^{4})$.

\section{Iterative method to Solve the Optimization Problem}

In order to demonstrate the benefits of our developed GM method for the UAV's optimal trajectory and Alice's transmit power design, in this section we employ the CI method to solve the optimization problem \eqref{mod:18} as a benchmark scheme.

This CI method applies the BCD to partition the variables into two blocks, i.e., the UAV's trajectory variable ${\bf{Q}}$ and the Alice's transmit power variable ${\bf{P}}$. Specifically, the CI method starts with determining the initial trajectory of the UAV ${\bf{Q}}^{(0)}$. Then, in each iteration $l$, this method first identifies ${\bf{P}}^{l}$ based on the trajectory ${\bf{Q}}^{l-1}$ obtained from the last iteration and designs ${\bf{Q}}^{l}$ based on the obtained ${\bf{P}}^{l}$, as shown in \eqref{mod:21A}. Since the optimization problem in each iteration is still non-convex, we adopt CCCP to handle it. We note that, however, there is no need to determine the UAV's initial trajectory in our developed GM method, which demonstrates the first benefits of the GM method.




\begin{align}\label{mod:21A}
&\underbrace {{{\bf{Q}}^{(0)}}}_{{\rm{Initialization}}} \to \underbrace {{{\bf{P}}^{(1)}} \to {{\bf{Q}}^{(1)}}}_{{\rm{First~ iteration}}} \to \ldots  \to \underbrace {{{\bf{P}}^{(l - 1)}} \to {{\bf{Q}}^{(l - 1)}}}_{{(l-1)\rm{-th iteration}}} \notag \\
&\to \underbrace {{{\bf{P}}^{(l)}} \to {{\bf{Q}}^{(l)}}}_{{(l) \rm{- th~ iteration}}} \to  \ldots  \to \underbrace {{{\bf{P}}^*} \to {{\bf{Q}}^*}}_{{(l-1)\rm{-th~ iteration}}}.
\end{align}

\subsection{Feasibility Analysis and Trajectory Initialization}
Based on  Lemma~\ref{lemma:3}, we know that there always exists a feasible $\mathbf{P}$ satisfying constraint \eqref{18b} for any feasible trajectory $\bf Q$ satisfying \eqref{18c}, \eqref{18d}, and \eqref{18e}. Thus, ${\bf Q}^{(0)}$ can be determined by only considering the trajectory constraints (i.e., \eqref{18c}, \eqref{18d}, and \eqref{18e}). Following \cite{Li[]}, we use the best-effort manner to initialize the trajectory ${\bf Q}^{(0)}$, where the UAV first tries its best to approach Willie and then returns to its ending position. For different values of $\mathbf{q}_{u,0}, \mathbf{q}_{u,F}$, and $\mathbf{q}_{w}, \mathbf{Q}^{(0)}$ can be obtained in the following two cases.

\begin{figure}[!t]
\centering
\includegraphics[width=0.87\textwidth]{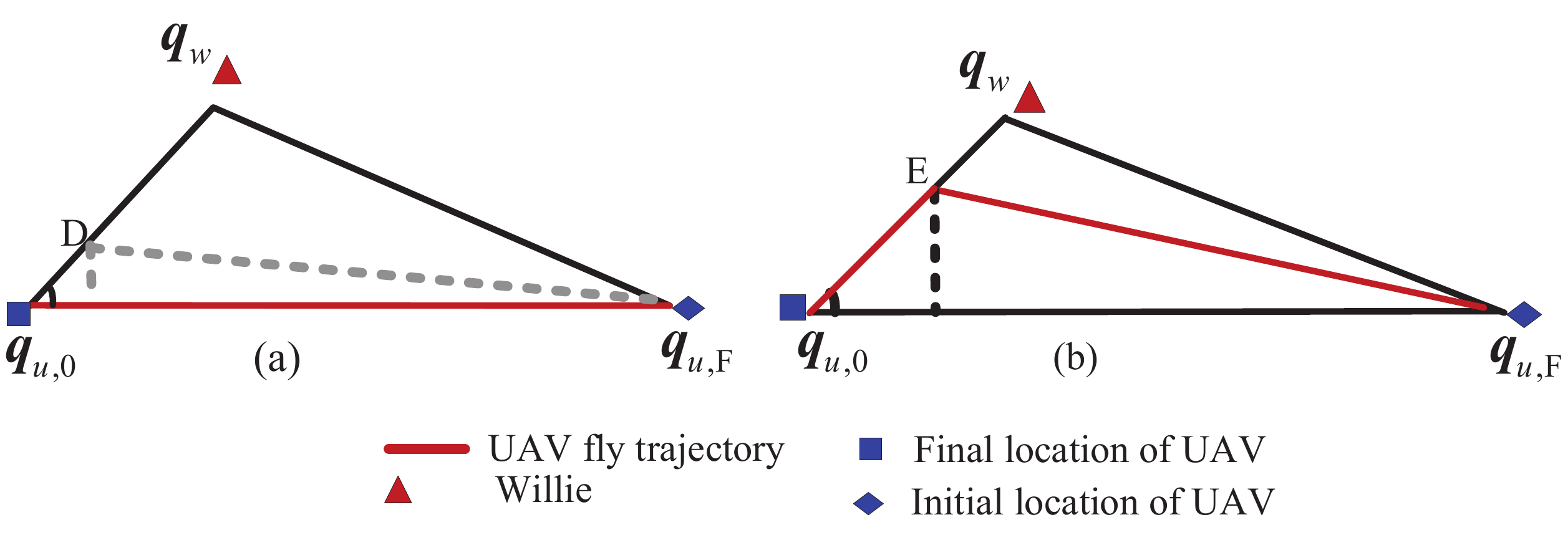}
\caption{The UAV's initial flight trajectory under difference cases.}
\vspace{-0.5cm}
\label{fig:3}
\end{figure}

$\mathbf{Case~1}$: $\left\lceil {\frac{||{{\mathbf{q}_w} - {\mathbf{q}_{u,0}}||}}{{{V_{\max }}{\delta _t}}}} \right\rceil + \left\lceil{\frac{||{{\mathbf{q}_{u,F}} - {\mathbf{q}_w}}||}{{{V_{\max }}{\delta _t}}}} \right\rceil \leqslant N$:

In this case, $N$ (i.e., the flight duration) is sufficiently large such that the UAV can first fly from the initial point to the vertical top of Willie and then fly to the final point.   The UAV's initial trajectory under this case is shown as the red line in Fig.~\ref{fig:3}(a). Thus, the UAV's corresponding trajectory under this case is given by

\begin{equation}\label{mod:t5}
\mathbf{q}_{u}[n]=
\begin{cases}
{\mathbf{q}_{u,0} + n{V_{\max }}{\delta _t}\theta_{2}},&{n < N_{2}}, \\
{{\mathbf{q}_w}}, &{N_{2}\leqslant n < N_{3}},\\
{\mathbf{q}_w}+(n-N_{3}){V_{\max }}{\delta _t}\theta_{3},& N_{3}\leqslant n < N,\\
\mathbf{q}_{u,F}, &n = N,
\end{cases}
\end{equation}
where $\theta_{2}=\frac{\mathbf{q}_{w} - \mathbf{q}_{u,0}}{||\mathbf{q}_{w} - \mathbf{q}_{u,0}||}$, $N_{2}=\left\lceil {\frac{||{{\mathbf{q}_w} - {\mathbf{q}_{u,0}}||}}{{{V_{\max }}{\delta _t}}}} \right\rceil$, $N_{3}=N - \left\lceil {\frac{||{{\mathbf{q}_{u,F}} - {\mathbf{q}_w}}||}{{{V_{\max }}{\delta _t}}}} \right\rceil$, and $\theta_{3}=\frac{\mathbf{q}_{u,F}-\mathbf{q}_{w}}{||\mathbf{q}_{u,F}-\mathbf{q}_{w}||}$.

\vspace{0.08cm}

$\mathbf{Case~2}$: $\left\lceil {\frac{||{{\mathbf{q}_w} - {\mathbf{q}_{u,0}}||}}{{{V_{\max }}{\delta _t}}}} \right\rceil + \left\lceil {\frac{||{{\mathbf{q}_{u,F}} - {\mathbf{q}_w}}||}{{{V_{\max }}{\delta _t}}}} \right\rceil >N$

In this case, $N$ is insufficient for the UAV to successively arrive at the vertical top of  Willie and then return to the final point. Thus, the initial flight trajectory of the UAV under this case is shown as the red line in Fig.~\ref{fig:3}(b), where the UAV first flies from the initial point to the direction of Willie and then turns to the direction of the final point at intermediate point ${\bf{q}}_{E}$ between the initial point and Willie. To proceed, we derive a closed-form expression for the intermediate point ${\bf{q}}_{E}$ in the following proposition.

\begin{proposition}\label{proposition1}
When $N$ is not sufficient for the UAV flying to the vertical top of Willie and then returning to its final point, the intermediate point ${\bf{q}}_{E}$ when the UAV turns its flying direction is given by
\begin{align} \label{mod:tt5}
\mathbf{q}_{E}=N_{4}V_{\max }{\delta _t}\theta_{2},
\end{align}
 where $N_{4}=\left\lfloor \frac{N^{2}(V_{\max}\delta_{t})^{2}-||{{\mathbf{q}_{u,F}} - {\mathbf{q}_{u,0}}}||^{2}}{2V_{\max}\delta_{t}(NV_{\max}\delta_{t}-||{{\mathbf{q}_{u,F}} - {\mathbf{q}_{u,0}}}||\cos(\theta_{2}-\theta_{3}))}\right\rfloor.$
\end{proposition}
\begin{IEEEproof}
The detailed proof is presented in Appendix D.
\end{IEEEproof}

 Then, following Proposition~\ref{proposition1} the UAV's initial trajectory under this case can be obtained as
\begin{equation}\label{mod:t7}
\mathbf{q}_{u}[n]=
\begin{cases}
{\mathbf{q}_{u,0} + n{V_{\max }}{\delta _t}\theta_{1}},&{n < N_{4}}, \\
{{\mathbf{q}_E} + (n-N_{4}){V_{\max }}{\delta _t}\theta_{3}},&
{N_{3} \leqslant n \leq N_{5}},\\
{\mathbf{q}_{u,F}},&N_{5}<n\leq N,
\end{cases}
\end{equation}
where $N_{5}=\left\lfloor||\frac{\mathbf{q}_{E}-\mathbf{q}_{u,F}||}{(V_{\max}\delta_{t})}\right\rfloor$ and $\theta_{4}=\frac{\mathbf{q}_{E}-\mathbf{q}_{u,F}}{||\mathbf{q}_{E}-\mathbf{q}_{u,F}||}$.

It should be noted that, when $N_{4}=0$ (i.e., the flight duration is only enough for the UAV to directly fly from the initial point to the final point) the intermediate point coincides with the initial point, i.e., $\mathbf{q}_{E}=\mathbf{q}_{u,0}$. Therefore, the UAV has to fly directly  from the initial point to the final point when $N_{4}=0$.

\subsection{Alice's Transmit Power Design}
For a given $\mathbf{Q}$, the optimization problem \eqref{mod:18} can be reformulated as
\begin{subequations}\label{mod:22}
\begin{align}
&\underset{\bf{P}}{\text{maximize }}\;\;\frac{1}{N}\sum\limits_{n = 1}^N {{{\log }_2}} \left( {1 + \psi[n] {P_a}\left[ n \right]} \right), \label{33a}\\
{\text{s.t. ~}}&\frac{{{P_a}\left[ n \right]\phi }}{\Gamma[n] } - \frac{{{P_a}\left[ n \right]\phi }}{\Gamma[n] }\exp \left( { - \frac{\Gamma[n] }{{{P_a}\left[ n \right]\phi }}} \right)\leq \epsilon,\label{33b}
\end{align}
\end{subequations}
where
\begin{equation}
\psi[n]  =  - \frac{{\ln (1 - {\rho _b}){\beta _0}d_{a,b}^{ - \alpha }}}{{{\hat{P}_u}[n]|{h_{u,b}}[n]{|^2} + \sigma _b^2}}.\notag
\end{equation}
We note that the left hand side (LHS) of \eqref{33b} is the difference of two convex functions, which renders the problem in a non-convex structure. To solve \eqref{mod:22}, we first exploit the CCCP method to deal with the LHS of the constraint \eqref{33b}.

We denote ${{\bf{P}}^{\left( l \right)}} = \left\{ {P_a^{\left( l \right)}\left[ n \right]},~\forall n \right\}$ as the fixed point of $P_a\left[ n \right]$ at the $l$-th iteration of the CCCP method. Then, the constraint \eqref{33b} can be rewritten as
\begin{equation}\label{mod:24}
\frac{{P_a^{\left( l \right)}[n]\phi }}{\Gamma[n] } - {{\rm{g}}_1}\left( {{P_a}[n]} \right) \le  \epsilon,
\end{equation}
where ${{\mathop{\rm g}\nolimits} _1}\left( {{P_a}[n]} \right)$ is obtained by approximating the second term of the LHS function of \eqref{33b} with its first order Taylor expansion, which is given by
\begin{align}\label{mod:25}
{{\rm{g}}_1}\left( {{P_a}[n]} \right){\rm{ }}& \buildrel \Delta \over = \frac{{P_a^{\left( l \right)}[n]\phi }}{\Gamma[n] }\exp \left( { - \frac{\Gamma[n] }{{P_a^{\left( l \right)}[n]\phi }}} \right)\\
 &+ \exp \left( { - \frac{\Gamma[n] }{{P_a^{\left( l \right)}[n]\phi }}} \right)\left( {\frac{ \phi}{{\Gamma[n] }} + \frac{1}{{P_a^{\left( l \right)}[n]}}} \right)\left( {{P_a}[n] - P_a^{\left( l \right)}[n]} \right)\nonumber.
\end{align}

We note that this approximation is generally with a high accuracy, since the transmit power in covert communications is normally in its low regime due to the covertness constraint. Such approximations have been widely used in the existing works on covert communications (e.g., \cite{Zhou1, Zhou2}). We also note that such an approximation is not required to determine the Alice's transmit power {\bf{P}} in our developed GM method in the last section. This could serve as one reason why the developed GM method can outperform the CI method.

Substituting \eqref{mod:25} into \eqref{mod:24}, ${\bf P}^{(l+1)}$ can be obtained by solving the following optimization problem
\begin{subequations}\label{mod:26}
\begin{align}
{{\bf{P}}^{\left( {l + 1} \right)}} = &\mathop {\arg \max }\limits_{\bf{P}} \;\;\frac{1}{N}\sum\limits_{n = 1}^N {{{\log }_2}} \left( {1 + \psi[n] {P_a}[n]} \right),\label{26a}\\
\text{s.t. ~~}
&\frac{{P_a^{\left( l \right)}[n]\phi }}{\Gamma[n] } - {{\rm{g}}_1}\left( {{P_a}[n]} \right) \le  \epsilon.\label{26b}
\end{align}
\end{subequations}

The above problem is a standard convex optimization problem, which can be efficiently solved by a convex optimization solver such as CVX \cite{Boyd}. The original optimization problem \eqref{mod:22} can be solved by solving \eqref{mod:26}
iteratively. At each iteration, the current optimal solution to \eqref{mod:26} gradually approaches
the optimal solution to \eqref{mod:22}.


%
%
%
%
%
%
%
%
%
%
%

\subsection{UAV's Trajectory Design}

After obtaining $ {\bf{P}} = \left\{ {{{P}_a}\left[ n \right],~\forall n} \right\}$ from \eqref{mod:26}, \eqref{mod:18} can be simplified as
\begin{subequations}\label{mod:37}
\begin{align}
\underset{\bf{Q}}{\text{maximize }}& \frac{1}{N}\sum\limits_{n = 1}^N {{{\log }_2}} \left( {1 - \frac{{\overline \psi[n]  }}{{\frac{{{{\hat P}_u}\left[ n \right]{\beta _0}}}{{{{\left\| {{{\bf{q}}_u}\left[ n \right] - {{\bf{q}}_b}} \right\|}^2} + {H^2}}} + \sigma _b^2}}} \right)\label{37a}\\
\text{s.t. ~~}&\overline {\mathop{\rm G}\nolimits}  \left( {{{\bf{q}}_u}\left[ n \right]} \right)\leq \epsilon, \forall n,\label{37b}\\
&\|\mathbf{q}_{u}[n]-\mathbf{q}_{u}[n-1]\|\leq V_{\max}\sigma_{t}, \forall n,\label{37c}\\
&\mathbf{q}_{u}[N]=\mathbf{q}_{u,F},\label{37d}\\
&\mathbf{q}_{u}[0]=\mathbf{q}_{u,0},\label{37e}
\end{align}
\end{subequations}
where $\bar \psi[n]=\ln (1 - {\rho _b}){ P_a}\left[ n \right]{\beta _0}d_{a,b}^{ - \alpha }$, and
\begin{align}
\overline {\mathop{\rm G}\nolimits}  \left( {{{\bf{q}}_u}\left[ n \right]} \right)& =\frac{{{{ P}_a}\left[ n \right]\left( {{{\left\| {{{\bf{q}}_u}\left[ n \right] - {{\bf{q}}_w}} \right\|}^2} + {H^2}} \right)}}{{{{\hat P}_u}[n]d_{a,w}^\alpha }}
 - \frac{{{{P}_a}\left[ n \right]\left( {{{\left\| {{{\bf{q}}_u}\left[ n \right] - {{\bf{q}}_w}} \right\|}^2} + {H^2}} \right)}}{{{{\hat P}_u}[n]d_{a,w}^\alpha \exp \left( {\frac{{{{\hat P}_u}[n]d_{a,w}^\alpha }}{{{{ P}_a}\left[ n \right]\left( {{{\left\| {{{\bf{q}}_u}\left[ n \right] - {{\bf{q}}_w}} \right\|}^2} + {H^2}} \right)}}} \right)}}. \notag
\end{align}

We note that the objective function is a concave function of ${\|\mathbf{q}_{u}[n]-\mathbf{q}_{b}\|^{2}}$, while the trajectory constraints \eqref{37c}, \eqref{37d} and \eqref{37e} are also convex sets of $\bf{Q}$. However, the covertness constraint \eqref{37b} is non-convex with respect to ${\|\mathbf{q}_{u}[n]-\mathbf{q}_{w}\|^{2}}$. Similar to solving the power control problem, we also use the CCCP method to deal with the non-convex covertness constraint.

We denote  ${{\bf{Q}}^{\left( l \right)}} = \left\{ {{\bf{q}}_u^{\left( l \right)}\left[ n \right],\forall n} \right\}$ as the fixed point of $\bf Q$ in the $l$-th iteration of the CCCP method.
Then,  \eqref{37b} can be transformed into
\begin{equation}\label{mod:28}
\frac{{\left( {{{\left\| {{\bf{q}}_u^{\left( l \right)}[n] - {{\bf{q}}_b}} \right\|}^2} + {H^2}} \right){{ P}_a}[n]}}{{{{\hat P}_u}[n]d_{a,w}^\alpha }} - {{\rm{g}}_2}\left( {{{\bf{q}}_u}[n]} \right) \le \epsilon,
\end{equation}
where
\begin{align}\label{mod:39}
{{\mathop{\rm g}\nolimits} _2}\left( {{{\bf{q}}_u}[n]} \right) \buildrel \Delta \over =& \exp \left( { - \frac{{{{\widehat P}_u}[n]d_{a,w}^\alpha }}{{{{ P}_a}[n]( {{{\| {{\bf{q}}_u^{( l )}[n] - {{\bf{q}}_w}} \|}^2} + {H^2}} )}}} \right)\frac{{{{ P}_a}[n]( {{{\| {{\bf{q}}_u^{( l )}[n] - {{\bf{q}}_w}} \|}^2} + {H^2}} )}}{{{{\hat P}_u}[n]d_{a,w}^\alpha }}\notag\\
&+ \frac{{\left( {\frac{{{P_a}[n]}}{{{{\hat P}_u}[n]d_{a,w}^\alpha }} + \frac{1}{{( {{{\| {{\bf{q}}_u^{( l )}[n] - {{\bf{q}}_w}} \|}^2} + {H^2}} )}}} \right)
( {{{\| {{{\bf{q}}_u}[n] - {{\bf{q}}_w}} \|}^2} - {{\| {{\bf{q}}_u^{( l )}[n] - {{\bf{q}}_w}} \|}^2}} )}}{{\exp \left( {\frac{{{{\hat P}_u}[n]d_{a,w}^\alpha }}{{{{ P}_a}[n]( {{{\| {{\bf{q}}_u^{( l )}[n] - {{\bf{q}}_w}} \|}^2} + {H^2}} )}}} \right)}},
\end{align}
which is obtained by approximating the second term of $\overline {\mathop{\rm G}\nolimits}  \left( {{{\bf{q}}_u}\left[ n \right]} \right)$ with its first order of Taylor expansion. Then, ${\bf Q}^{(l)}$ can be obtained by solving the following optimization problem
\begin{subequations}\label{mod:38}
\begin{align}
&{{\bf{Q}}^{\left( {l + 1} \right)}}= \mathop {\arg \max }\limits_{\bf{Q}} \frac{1}{N}\sum\limits_{n = 1}^N {\log } \left( {1 - \frac{{\bar \psi[n] }}{{\frac{{{{\hat P}_u}\left[ n \right]{\beta _0}}}{{{{\left\| {{{\bf{q}}_u}\left[ n \right] - {{\bf{q}}_b}} \right\|}^2} + {H^2}}} + \sigma _b^2}}} \right)\\
\text{s.t.~}&\frac{{\left( {{{\left\| {{\bf{q}}_u^{\left( l+1 \right)}[n] - {{\bf{q}}_b}} \right\|}^2} + {H^2}} \right){{ P}_a}[n]}}{{{{\hat P}_u}[n]d_{a,w}^\alpha }} - {{\rm{g}}_2}\left( {{{\bf{q}}_u}[n]} \right) \le \epsilon, \forall n,\label{38b}\\
&\|\mathbf{q}_{u}^{(l+1)}[n+1]-\mathbf{q}_{u}^{(l+1)}[n]\|\leq V_{\max}\sigma_{t},\label{38c}\\
&\mathbf{q}_{u}^{(l+1)}[N]=\mathbf{q}_{u,F}^{(l+1)},\label{38d}\\
&\mathbf{q}_{u}^{(l+1)}[0]=\mathbf{q}_{u}^{(l+1)}[0].\label{38e}
\end{align}
\end{subequations}
The objective function and constraints of the above optimization problem are all in convex structures. As such, \eqref{mod:38} is a standard convex optimization problem and can be effectively solved by CVX \cite{Boyd}. The original optimization problem \eqref{mod:37} can be solved
by iteratively solving \eqref{mod:38} until achieving the convergence.

\subsection{Overall Algorithm for the CI Method}

Based on the solutions to the sub-problems \eqref{mod:26} and \eqref{mod:38}, we can apply the CI method to solve the optimization problem \eqref{mod:18}. The overall procedure is summarized in {\bf Algorithm 2}.

\begin{algorithm}[htb]

\caption{CI method for Trajectory and Power Optimization}             

\label{alg:Framwork}                  

\begin{algorithmic}[1]                

\STATE $\mathbf{Initialize}$:

\STATE Set $l=0$, tolerance $\sigma>0$;

\STATE Initialize $Q^{(0)}$ based $\eqref{mod:t5}$ or $\eqref{mod:t7}$;

\STATE ${\bf repeat}$: \label{step:2}

\STATE $l=l+1$;

\STATE For given feasible $\{\mathbf{Q}^{(l-1)}, \mathbf{P}^{(l-1)}\}$, solve problem \eqref{mod:26} and  obtain the current optimal solution
 $\mathbf{P}^{(l)}$;

\STATE Given feasible $\mathbf{P}^{(l)}$, ${\mathbf{Q}^{(l)}}$ is obtained by the \eqref{mod:38};

\STATE $\mathbf{Until:}$ $|\mathcal{L}\left(\mathbf{Q}^{(l)},\mathbf{P}^{(l)}\right)-\mathcal{L}\left(\mathbf{Q}^{(l-1)},\mathbf{P}^{(l-1)}\right)|\leq \sigma$;\label{step:5}

\STATE  Set $\left\{\mathbf{Q}^{\ast},\mathbf{P}^{\ast}\right\}=\left\{\mathbf{Q}^{(l)},\mathbf{P}^{(l)}\right\};$


\STATE $\mathbf{Return}$: $\left\{\mathbf{Q}^{\ast},\mathbf{P}^{\ast}\right\}$. \label{step:7}
\end{algorithmic}
\end{algorithm}


We note that {\bf Algorithm 2} is not guaranteed to converge to a global optimal point, since only a local optimal solution to the UAV's power control or trajectory design can be determined in each iteration. In addition, we note that the three convex restriction formulations \eqref{26b}, \eqref{38b} and \eqref{38c} involve LMI and second-order cone (SOC) constraints. As such, the overall complexity of {\bf Algorithm 2} is on the order of $\mathcal{O}(N^{\frac{9}{2}})$ as per the analyses in \cite{KYWang}. We recall that the complexity of {\bf Algorithm 1} for our developed GM method is on the order $\mathcal{O}(N^{4})$ as detailed in the last section. Therefore, comparing with the CI method, we note that the GM method is of a lower complexity.

\section{Numerical Results}

In this section, numerical results are presented to evaluate the performance of the UAV-aided covert communication with the UAV's optimal trajectory and Alice's transmit power achieved by our proposed GM method. To demonstrate the benefit of our developed GM design, we compare it with the benchmark CI method as detailed in Section IV. The simulation parameters are set as: $q_{0}=[-100\text{m},100\text{m}]^{T}$, $q_{F}=[500\text{m},100\text{m}]^{T}$, $H=100$ m, $V=3$ m/s, $\sigma_{t}=0.5$ s, $q_{a}=[0,0]^{T}$, $q_{b}=[200\text{m},0]^{T}$, $q_{w}=[200\text{m},200\text{m}]^{T}$, $\rho_{b}=0.1$, $\beta_{0}=-60$ dB, and $\frac{\beta_{0}}{\sigma^{2}}=80$ dB.

\begin{figure}[!h]
\centering
\includegraphics[width=1\textwidth]{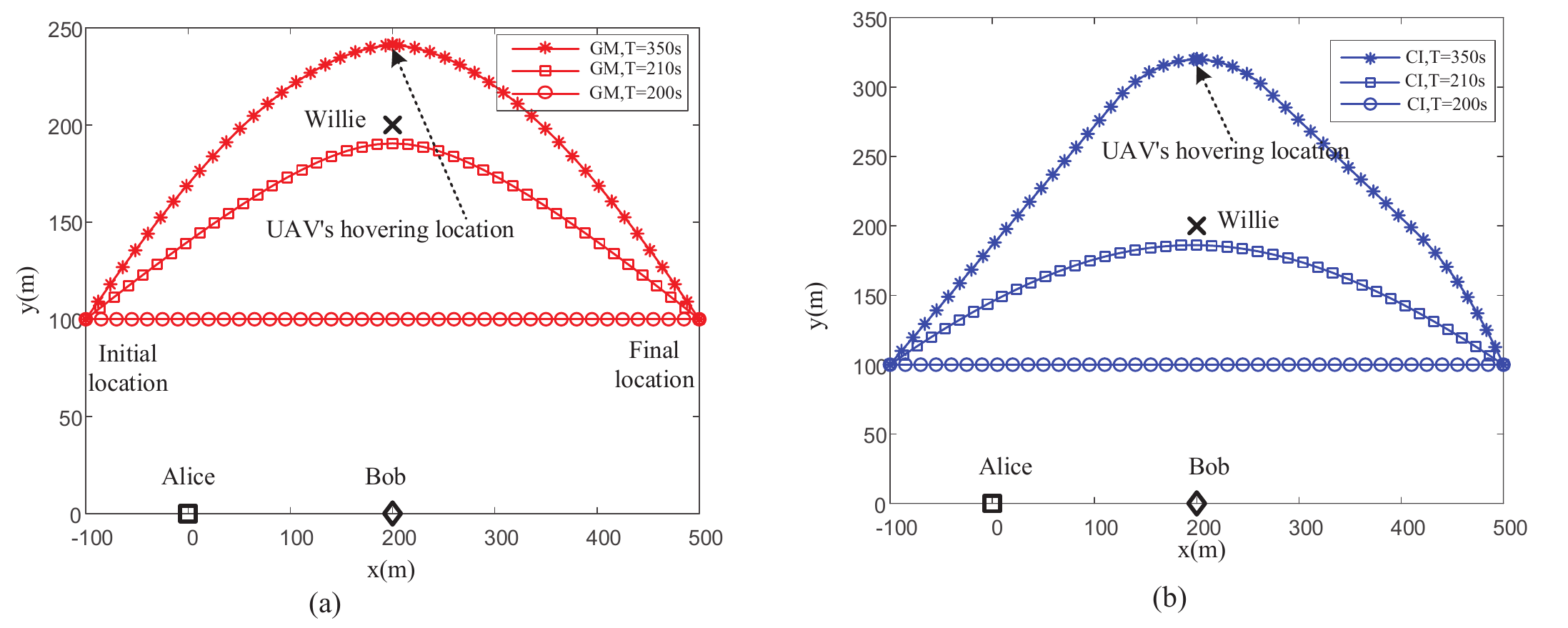}
\caption{The UAV's optimal trajectories for different values of the flight period $T$ with $\epsilon=0.1$, where (a) is obtained by the GM method and (b) is obtained by the CI method.}
\label{fig:13}
\end{figure}

\begin{figure}[!t]
\centering
\includegraphics[width=0.7\textwidth]{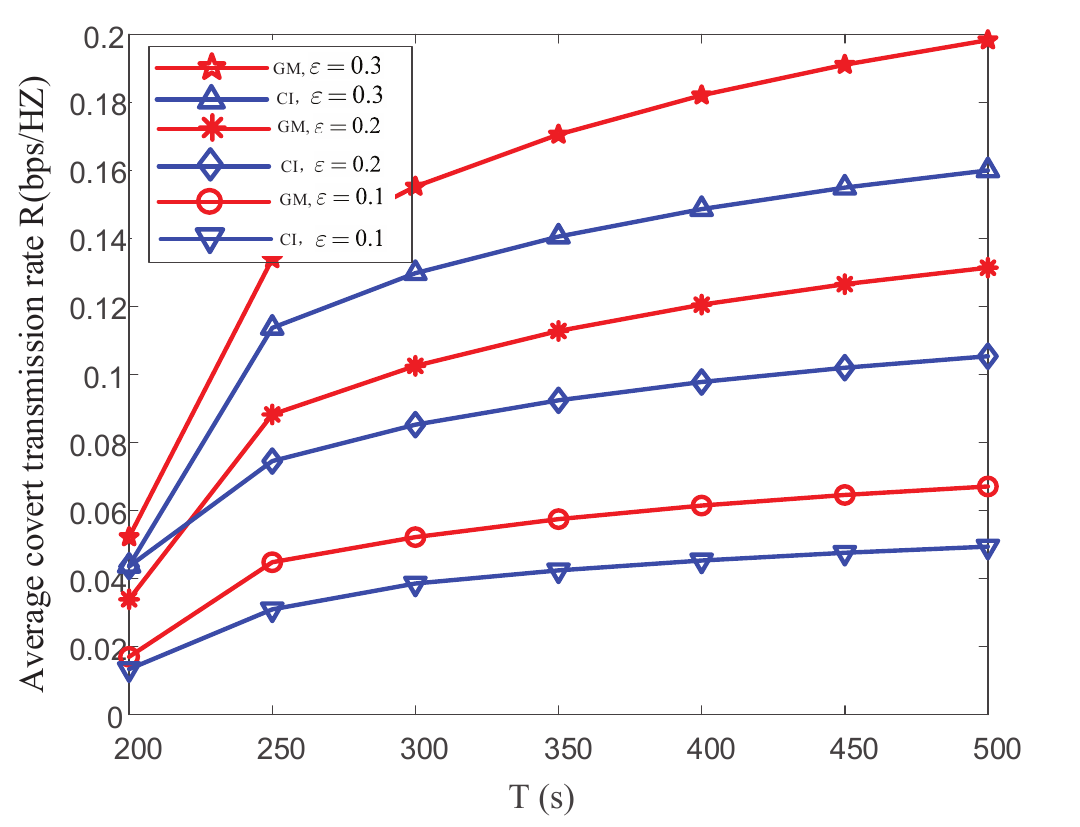}
\caption{Average covert rate achieved by the GM and CI methods versus of the UAV's flight period $T$ for different values of covertness parameter $\epsilon$.}
\label{fig:17}
\end{figure}

In Fig.~\ref{fig:13}(a) and Fig.~\ref{fig:13}(b), we plot the optimal trajectories of the UAV achieved by our developed GM method and the benchmark CI method with different values of the flight period $T$, respectively, where  Alice, Bob, Willie, and the UAV's
initial and final locations are marked with $\Box$, $\lozenge$, $\times$, and $\bigtriangleup$, respectively.
In Fig.~\ref{fig:13}(a), we first observe that the trajectory achieved
by our developed GM method always shrinks inward closer to Willie relative to that achieved by the CI method as shown in Fig.~\ref{fig:13}(b). This could serve as one reason why our proposed GM method can outperform the benchmark CI method, in terms of achieving a higher average covert rate as shown in Fig.~\ref{fig:17}. This observation is mainly due to the fact that the CI method uses a large number of approximations, e.g., \eqref{mod:39}, in solving the related optimization problem. We recall that the complexity of our developed GM method is also lower than that of the CI method, which is another advantage of the GM method.


\begin{figure}[!t]
\centering
\subfigure[]{
\begin{minipage}{7.4cm}
\includegraphics[width=1\textwidth]{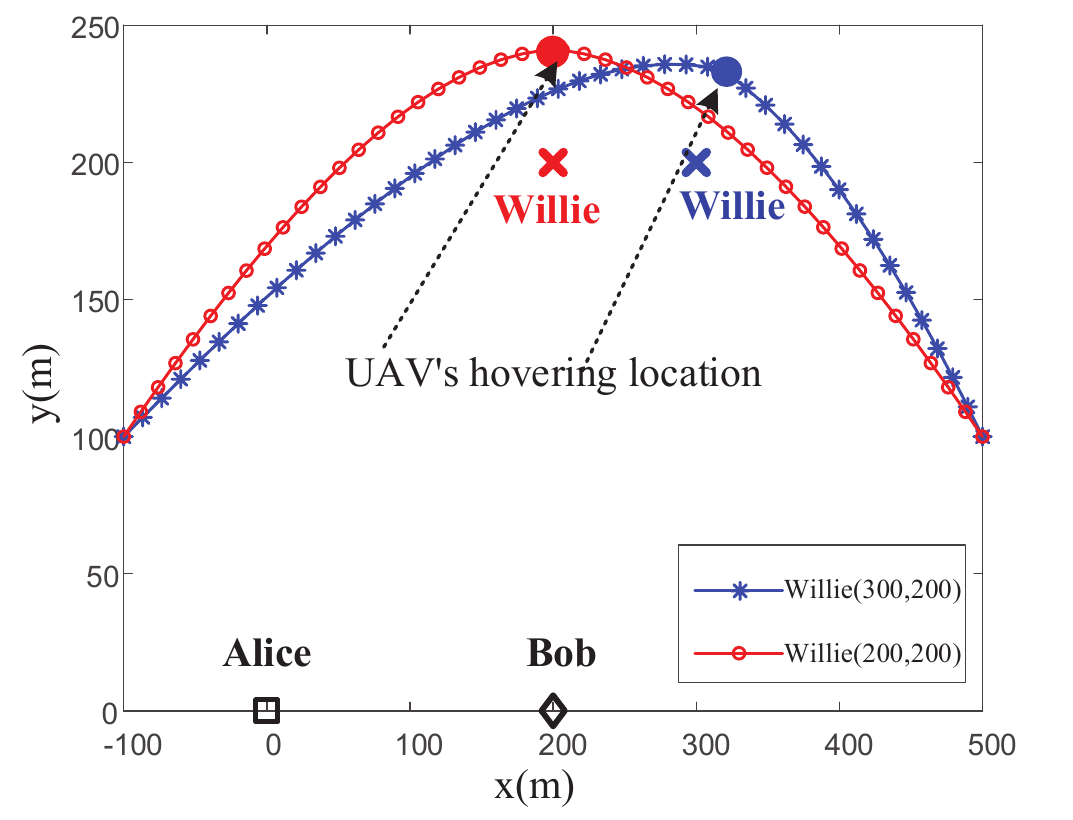}
\end{minipage}}
\hspace{2em}
\subfigure[]{
\begin{minipage}{7.4cm}
\includegraphics[width=1\textwidth]{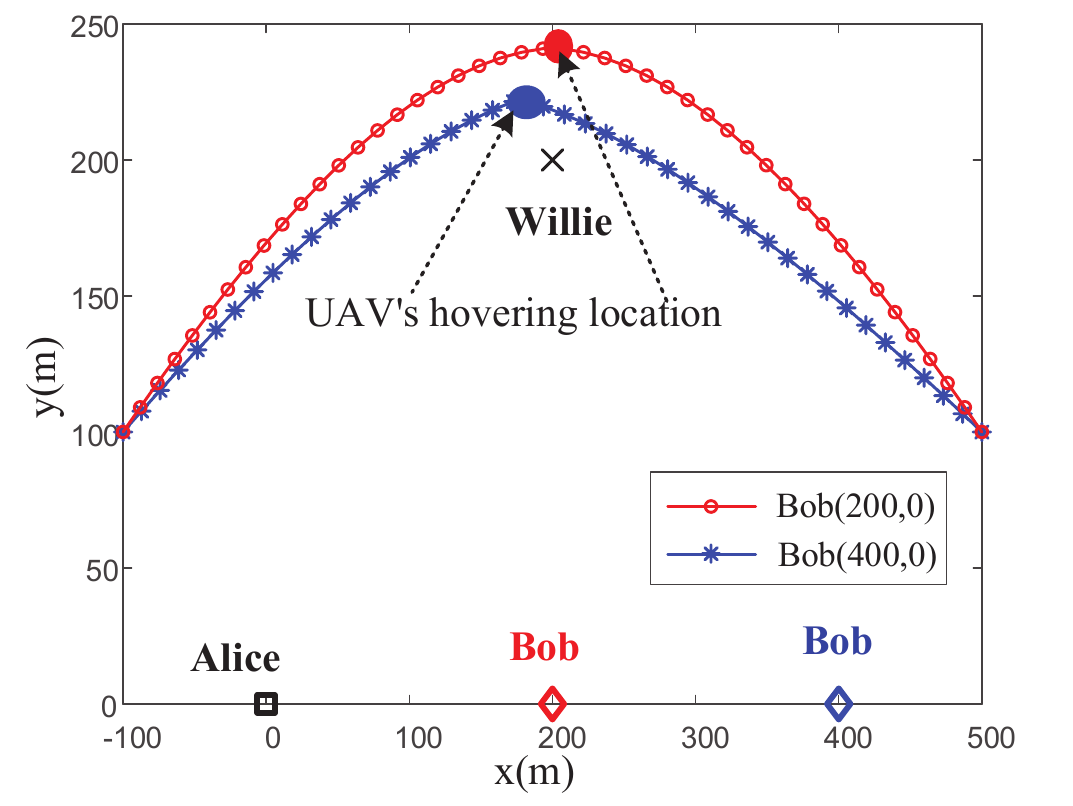}
\end{minipage}}
\caption{The UAV's optimal trajectory achieved by our developed GM method for different locations of Willie in (a) and different locations of Bob in (b), where $\epsilon=0.1$ and $T=350$s.}
\label{fig:13A}
\end{figure}

In Fig.~\ref{fig:13A}(a) and Fig.~\ref{fig:13A}(b), we plot the UAV's optimal trajectories achieved by our developed GM method for different locations of Willie and Bob, respectively. In Fig.~\ref{fig:13A}(a), as expected we observe that the optimal trajectory of the UAV changes with Willie's location. This is mainly due to the fact that Willie's location affects his detection performance of Alice's transmission, which leads to different optimal trajectories of the UAV to satisfy the covertness constraint. In addition,
the UAV needs to make a trade-off between hiding Alice's communication by creating interference to Willie and avoiding interference to Bob. In addition, we observe that the UAV's hovering location (shown as filled circles in this figure) is on the ray from Bob to Willie, which is consistent with maximizing the objective function in the optimization problem \eqref{mod:ss2} subject to some constraints. As expected, in Fig.~\ref{fig:13A}(b) we observe that the UAV's optimal trajectory varies with Bob's location. As explained above, this is due to the fact that the UAV has to balance between avoiding interference to Bob and creating interference to Willie. Specifically, when Bob moves further away from Alice, to maintain a similar covert rate Alice has to increase her transmit power, which requires more interference created by the UAV to Willie in order to meet the same covertness level (i.e., guaranteeing the same detection performance at Willie). Again, in this figure we also observe that the UAV's optimal hovering location is on the ray from Bob to Willie, which is consistent with the observation made in Fig.~\ref{fig:13A}(a).

\begin{figure}[!t]
\centering
\includegraphics[width=0.7\textwidth]{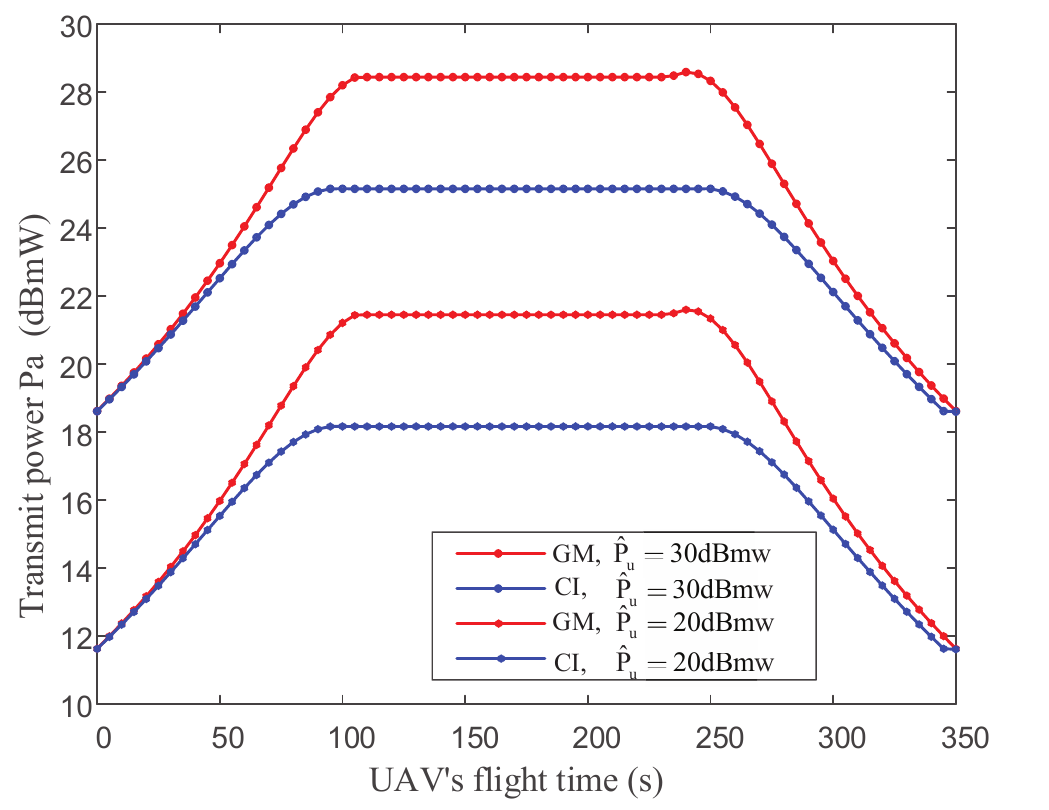}
\caption{Alice's optimal transmit power versus the UAV's flight time for different values of the UAV's maximum jamming transmit power $\hat{P}_{u}$ by UAV, where $\epsilon=0.1$ and $T=350$s.}
\label{fig:15}
\end{figure}

In Fig.~\ref{fig:15}, we plot Alice's optimal transmit power achieved by our GM method and the benchmark CI method versus the UAV's flight time for different values of the UAV's maximum jamming power ${\hat P}_{u}\left[ n \right]$. In this figure, we first observe that Alice's optimal transmit power $P_{a}[n]$  achieved
by the GM method is larger than that achieved by the CI method. This is due to the fact that the UAV's trajectory achieved by the GM method is always closer to Willie than that achieved by the CI method, as shown in Fig.~\ref{fig:13}. This leads to more interference from the UAV to Willie in the design based on the GM method, which enables Alice to transmit with a higher power while satisfying the covertness constraint. We also observe that Alice's optimal
transmit power $P_{a}[n]$ increases with the UAV's maximum transmit power ${\hat P}_{u}\left[ n \right]$. This is due to the fact that the equality in the covertness constraint always holds in the solution (i.e., $P_{a}[n]\tau[n]\left(1-e^{-\frac{1}{P_{a}[n]\tau[n]}}\right)= \epsilon$), while $\tau[n]$ monotonically decreases with ${\hat P}_{u}\left[ n \right]$ and $P_{a}[n]\tau[n]\left(1-e^{-\frac{1}{P_{a}[n]\tau[n]}}\right)$ increases with $P_{a}[n]$. Furthermore, we observe that Alice' s optimal transmit power $P_{a}[n]$ first increases, then stays at a constant level, and finally decreases as the flight time increases. The main reason is that the interference to Willie increases as the UAV moves closer to Willie (which enables Alice to transmit with a higher power) and such interference stays at a relative stable level when the UAV is hovering around Willie (which makes Alice to transmit with a constant power).


In Fig.~\ref{fig:17}, we plot the average covert rate achieved by our developed GM method  and the benchmark CI method versus the UAV's flight period $T$ for different values of  $\epsilon$ (representing different required covertness levels). As expected from the observations made in Fig.~\ref{fig:13} and Fig.~\ref{fig:15}, in this figure we first observe that our developed GM method outperforms the benchmark CI method in terms of achieving a significantly higher average covert rate. We recall that this is mainly due to the fact that the UAV's optimal trajectory achieved by the GM method always shrinks inward closer to Willie than that achieved by the CI method and thus the UAV can create more interference to Willie in the GM method relative to in the CI method, which enables Alice to transmit with higher power while still satisfying the covertness constraint. We also recall that, as analyzed in Sections III and IV, the complexity of the GM method is lower than that of the CI method. Therefore, this observation confirms the superiority of the developed GM method for the joint UAV's trajectory and Alice's transmit power design to achieve covert communications. In this figure, we also observe that the average covert rate slightly increases with the UAV's flight period $T$, since the UAV has relatively more time to hover nearby Willie as $T$ increases.

\section{Conclusions}
In this work, for the first time, we developed the GM method to solve the joint optimization problem of the
UAV's trajectory and Alice's transmit power in the context of UAV-aided covert communications. The considered optimal design was to maximize the average covert transmission rate subject to the transmission outage and covertness constraints. The GM method did not require approximations based on Taylor expansions or the UAV's trajectory initialization. In the GM method, the UAV's optimal trajectory was determined by efficiently solving an equation set, based on which Alice's optimal transmit power was analytically derived. In order to demonstrate the benefits of our developed GM method, we also detailed the CI method as a benchmark to solve the joint optimization problem. Compared with the CI method, the GM method achieved better covert communication performance with a lower computational complexity.

\appendices

{\section{Proof of lemma~\ref{lemma:3}\label{app:lemma:3}}
\begin{IEEEproof}
We define $f(x)=x\left(1-e^{-\frac{1}{x}}\right)$, where $x>0$. We first determine the first-order derivative of $f(x)$ with respect to $x$, which is given by
\begin{equation}\label{mod:58}
f'(x)=1-e^{-\frac{1}{x}}(1+\frac{1}{x}).
\end{equation}
Considering $x\geq 0$ and that $x=0$ is the singularity of $f'(x)$, we cannot determine the sign of $f'(x)$ for $x=0$. To proceed, we derive the second-order derivative of $f(x)$ with respect to $x$ as
\begin{equation}\label{mod:581}
f''(x)=-\frac{1}{x^{3}}e^{-\frac{1}{x}}<0,
\end{equation}
which indicates that the first derivative of $f(x)$, i.e., $f'(x)$, is a monotonically decreasing function of $x$. As per \eqref{mod:58}, we also have
\begin{equation}\label{mod:582}
\begin{aligned}
\lim \limits_{x\to +\infty}{f'(x)}&=\lim \limits_{x\to +\infty}(1-e^{-\frac{1}{x}}(1+\frac{1}{x}))\\
&=\lim \limits_{x\to +\infty}1-\lim \limits_{x\to +\infty}e^{-\frac{1}{x}}(1+\frac{1}{x})\\
&=1-\lim \limits_{x\to +\infty}\frac{1}{x^{3}e^{\frac{1}{x}}}\\
&=1,
\end{aligned}
\end{equation}
which, together with \eqref{mod:581}, proves that $f(x)$ is an increasing function of $x$. As such, replacing ${P_a}[n]$ with $x$ in \eqref{mod:582}, we can obtain the ${\rm G}\left( {{P_a}\left[ n \right],{{\bf{q}}_u}\left[ n \right]} \right)$ in \eqref{s20} is a monotonically increasing function of ${P_a}[n]$.
\end{IEEEproof}}

{\section{Proof of theorem~\ref{theorem:4}\label{app:theorem:4}}
\begin{IEEEproof}
We define $\overline {\bf{q}} [n] = {\left[ {{x_n},{y_n},{z_n}} \right]^T}$, which satisfies $\frac{{\left\| {\overline {\bf{q}} [n] - {{\bf{q}}_b}} \right\|}}{{\left\| {\overline {\bf{q}} [n] - {{\bf{q}}_w}} \right\|}} = k$. We recall that the coordinates of Bob (B) and Willie (W) are $\mathbf{q}_{b}=(x_{b},y_{b},0)$ and $\mathbf{q}_{w}=(x_{w},y_{w},0)$, respectively.

When $k\neq1$, as per $\frac{||\mathbf{\bar{q}}[n]-\mathbf{q}_{b}||^{2}}{||\mathbf{\bar{q}}[n]-\mathbf{q}_{w}||^{2}}=k^{2}$, we have
\begin{equation}\label{s44}
\begin{aligned}
(x_{n}-x_{b})^{2}+(y_{n}-y_{b})^{2}+(z_{n})^{2}=
k^{2}((x_{n}-x_{w})^{2}+(y_{n}-y_{w})^{2}+(z_{n})^{2}).
\end{aligned}
\end{equation}
After some  mathematical simplifications, \eqref{s44} can be rewritten as
\begin{equation}\label{s45}
\begin{aligned}
\left(x_{n}+x'\right)^{2}+\left(y_{n}+y'\right)^{2}+(z_{n})^{2}=\left(c\frac{k}{k^{2}-1}\right)^{2},
\end{aligned}
\end{equation}
where $x'=\frac{k^{2}x_{w}-x_{b}}{1-k^{2}}$, $y'=\frac{k^{2}y_{w}-x_{b}}{1-k^{2}}$ and $c= \|\mathbf{q}_{b}-\mathbf{q}_{w}\|^{2}$. We note that \eqref{s45} is the equation for a standard sphere whose center and radius are $(-\frac{k^{2}x_{w}-x_{b}}{1-k^{2}},-\frac{k^{2}y_{w}-x_{b}}{1-k^{2}},0)$ and $\frac{ck}{k^{2}-1}$, respectively.

When $k=1$, as per $\frac{||\mathbf{\bar{q}}[n]-\mathbf{q}_{b}||^{2}}{||\mathbf{\bar{q}}[n]-\mathbf{q}_{w}||^{2}}=1$, we have
\begin{equation}\label{s46}
\begin{aligned}
(x_{n}-x_{b})^{2}+(y_{n}-y_{b})^{2}+(z_{n})^{2}=
((x_{n}-x_{w})^{2}+(y_{n}-y_{w})^{2}+(z_{n})^{2}),
\end{aligned}
\end{equation}
which can be simplified as
\begin{equation}\label{s47}
\begin{aligned}
x_{n}(x_{w}-x_{b})+y_{n}(y_{w}-y_{b})=0.
\end{aligned}
\end{equation}
We note that \eqref{s47} determines a line connected Bob and Willie.
Combining the results in \eqref{s45} and \eqref{s47}, this completes the proof of Theorem \ref{theorem:4}.
\end{IEEEproof}}

{\section{Proof of theorem~\ref{theorem:5}\label{app:theorem:5}}
\begin{IEEEproof}
We prove this theorem in two steps, where in the first step we determine the value range of $k$ the feasible ${\bf q}_u[n-1]$ and in the second step we derive the optimal $k$.



\begin{figure}[!t]
\centering
\subfigure[]{
\begin{minipage}{3.1cm}
\includegraphics[width=2.1\textwidth]{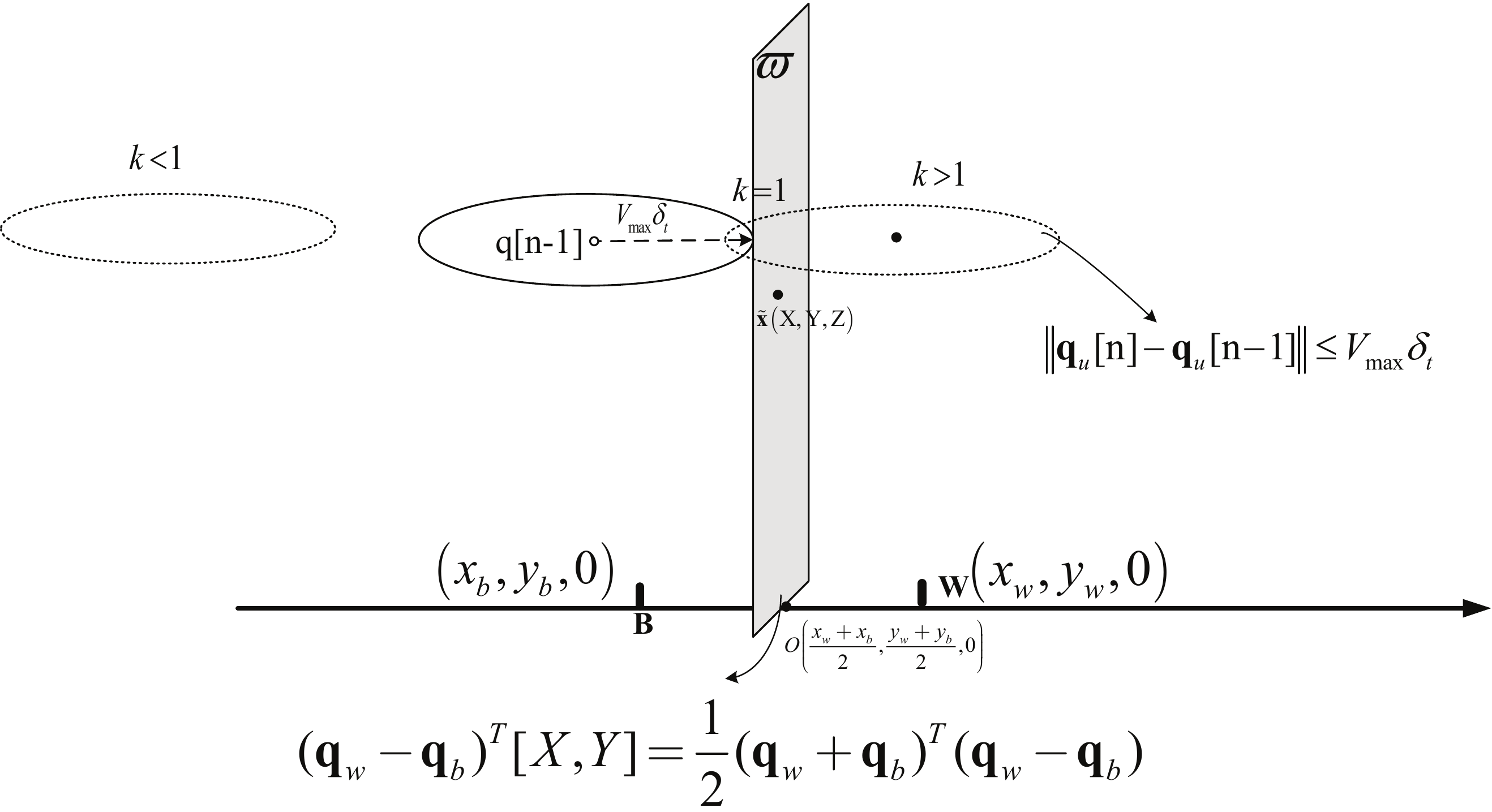}
\end{minipage}}
\hspace{8em}
\subfigure[]{
\begin{minipage}{2.7cm}
\includegraphics[width=2\textwidth]{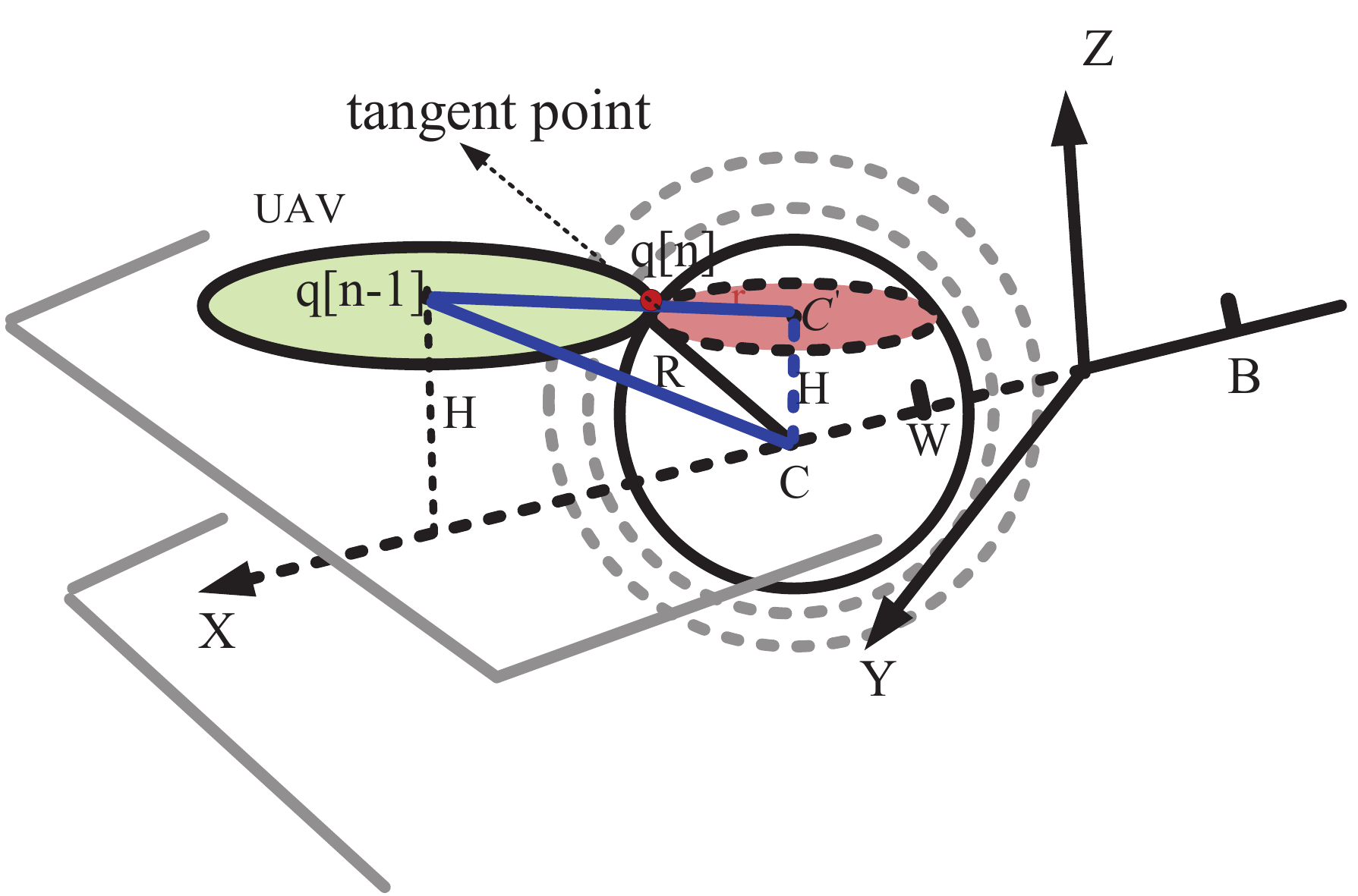}
\end{minipage}}
\hspace{2em}

\subfigure[]{
\begin{minipage}{2.9cm}
\includegraphics[width=2.15\textwidth]{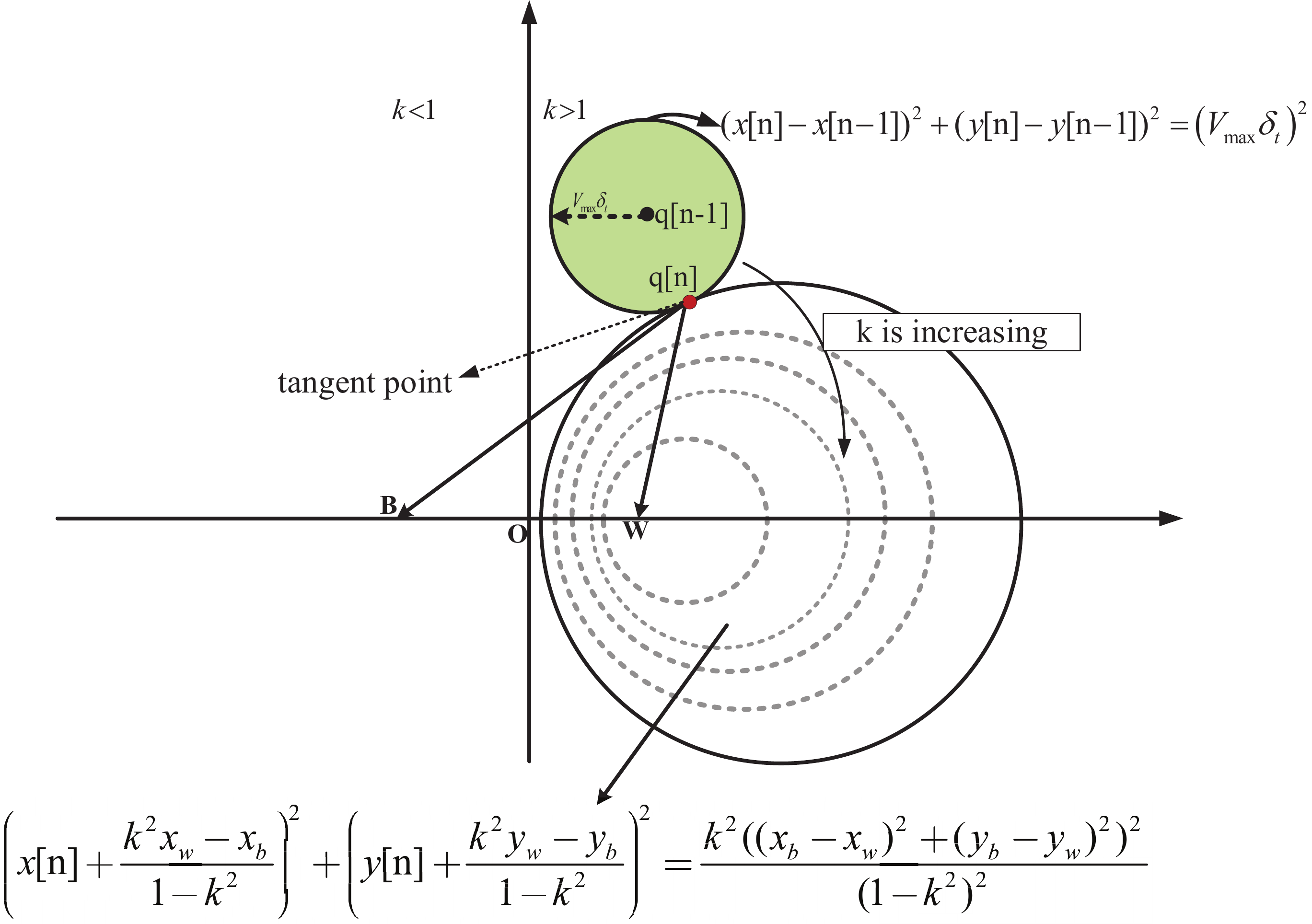}
\end{minipage}}
\hspace{7em}
\subfigure[]{
\begin{minipage}{2.9cm}
\includegraphics[width=2\textwidth]{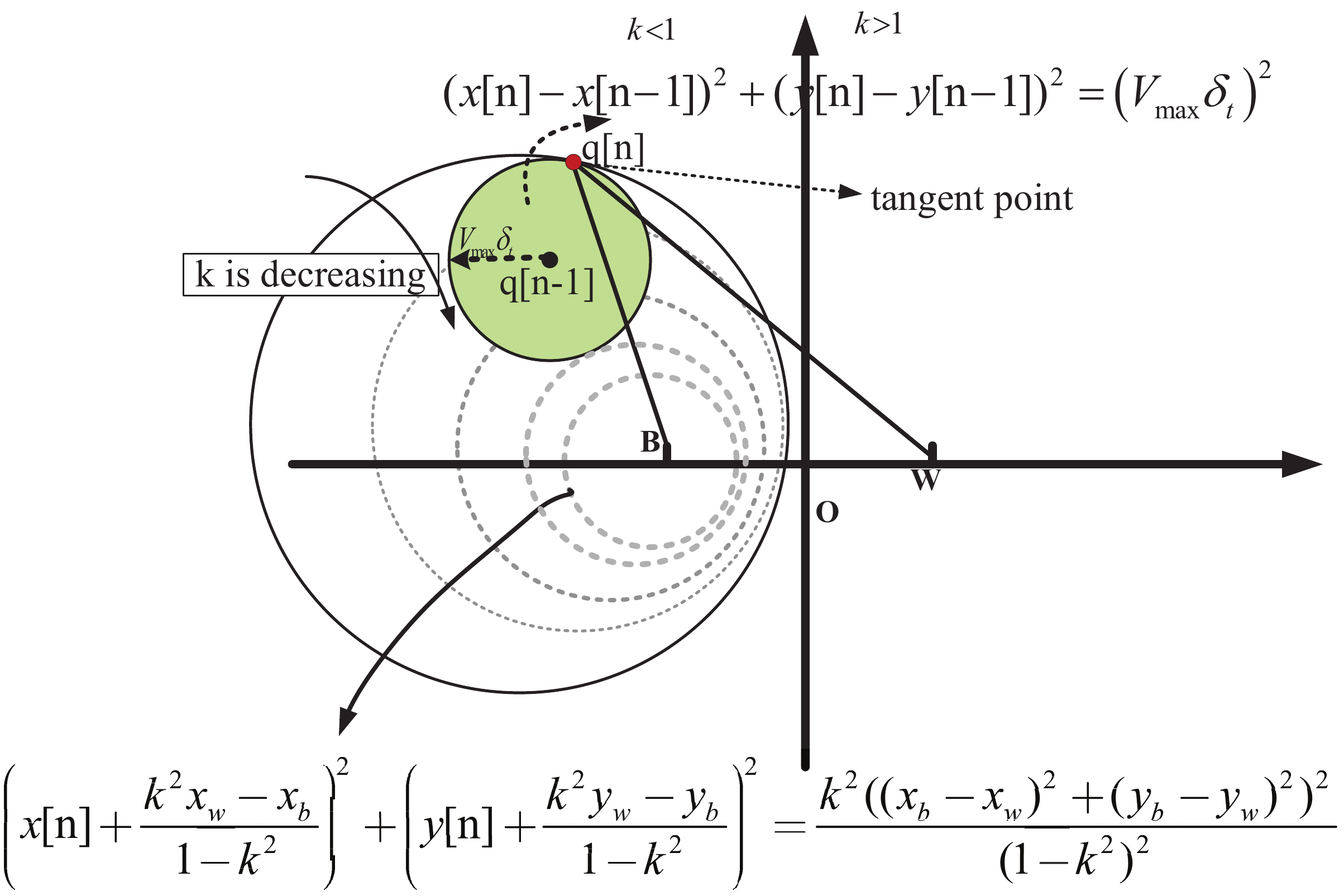}
\end{minipage}}
\caption{(a) Classification of $k$ values. (b) The best trajectory position of UAV in the next time slot.
(c) Radius change process when $k>1$. (d) The changing process of the radius of the Apollonius of Sphere when $k<1$.}
\label{fig:11}
\end{figure}


As the flight of UAV is a dynamic process, we know that the feasible region where the UAV can fly is constantly changing in each time slot.
We first define $\mathbf{O}=\frac{\mathbf{q}_{b}+\mathbf{q}_{w}}{2}$ and the plane perpendicular to $\overrightarrow{\mathbf{BW}}$ is denoted as $\varpi$, as shown in the gray rectangular area in Figure.~\ref{fig:11}(a). Then, we denote an arbitrary point in plane $\varpi$ as $\mathbf{\tilde{x}}=(X,Y,Z)$. Using the  property of the straight line perpendicular to the plane, we have
\begin{equation}\label{mod:81}
\begin{aligned}
&\overrightarrow{\mathbf{BW}}\cdot \overrightarrow{\mathbf{O\tilde{x}}}=0.
\end{aligned}
\end{equation}
After some mathematical simplifications, \eqref{mod:81} can be rewritten as
\begin{equation}\label{mod:82}
\begin{aligned}
(\mathbf{q}_{w}-\mathbf{q}_{b})^{T}[X,Y]=\frac{1}{2}(\mathbf{q}_{w}+\mathbf{q}_{b})^{T}(\mathbf{q}_{w}-\mathbf{q}_{b}).
\end{aligned}
\end{equation}
According to the constraint \eqref{24b}, the trajectory of the next time slot of the UAV needs to be within the feasible region. Therefore, we need to determine the position of the feasible region of the UAV at this time by the shortest distance from the $\mathbf{q}_{u}[n-1]$ to the plane $\varpi$, and which can be obtained by solving the optimization problem given by
\begin{equation}\label{mod:83}
\begin{aligned}
&\mathop {\min }\limits_{\mathbf{\tilde{x}}}\|\mathbf{q}_{u}[n-1]-\mathbf{\tilde{x}}\|^{2} \\ \nonumber
&\text{s.t. ~}
I[X,Y]=\hat{b}.
\end{aligned}
\end{equation}
By using the Lagrange Multiplier Method, the optimal $\mathbf{\tilde{x}}$ in the above optimization problem can be determined as
\begin{equation}\label{mod:84}
\begin{aligned}
\mathbf{\tilde{x}}=\mathbf{q}_{u}[n-1]-I^{T}(II^{T})^{-1}(I\mathbf{q}_{u}[n-1]-\hat{b}),
\end{aligned}
\end{equation}
where $I=(\mathbf{q}_{w}-\mathbf{q}_{b})^{T}$ and $\hat{b}=\frac{1}{2}(\mathbf{q}_{w}+\mathbf{q}_{b})^{T}(\mathbf{q}_{w}-\mathbf{q}_{b})$.
Combining \eqref{mod:81}, \eqref{mod:82} and \eqref{mod:83}, we have the following three cases regarding the value of $k$ as shown in Figure.~\ref{fig:11}(a):
\begin{itemize}
  \item We have $k = 1$ and $||\mathbf{q}_{u}[n]-\mathbf{q}_{b}||=||\mathbf{q}_{u}[n]-\mathbf{q}_{w}||$ for $\|\mathbf{q}_{u}[n-1]-\mathbf{\tilde{x}}\|=V_{\max}\sigma_{t}$.
  \item We have $k<1$ and $||\mathbf{q}_{u}[n]-\mathbf{q}_{b}||\leq||\mathbf{q}_{u}[n]-\mathbf{q}_{w}||$ for $\|\mathbf{q}_{u}[n-1]-\mathbf{\tilde{x}}\|> V_{\max}\sigma_{t}$.
  \item We have $k\geq1$ otherwise.
\end{itemize}


We next prove that the tangent point between the UAV's current location and the sphere determined by \eqref{s45} is the best location for the UAV in the next time slot, as shown in Figure.~\ref{fig:11}(b). In the following, we derive the optimal values of $k$ for the cases with $k<1$ and $k\geq 1$. To this end, we note that, when $k\neq 1$ the radius of the sphere determined by \eqref{s45} is $R(k)=\frac{ck}{k^{2}-1}$.

For $k>1$, the first derivative of $R(k)$ with respect to $k$ is given by
\begin{equation}\label{mod:c88}
\begin{aligned}
(R(k))'=\left(\frac{ck}{k^{2}-1}\right)'=-\frac{c(k^{2}+1)}{(k^{2}-1)^{2}}<0.
\end{aligned}
\end{equation}
As such, we note that $R(k)$ is a monotonically decreasing function of $k$. In order to provide some intuitive understanding, we provide the top view of how the radius changes with $k$ for $k > 1$ in Figure.~\ref{fig:11}(c).


From Figure.~\ref{fig:11}(c), we know that as $R(k)$ decreases, the Apollonius of Sphere moves further away from the feasible flight region of the UAV. In addition, the sphere center coordinate is given by $\left(\frac{k^2x_{w}-x_{b}}{k^2-1},\frac{k^2y_{w}-x_{b}}{k^2-1},0\right)$. Thus, as $k\rightarrow\infty$, the sphere center approaches to the point $\mathbf{q}_{w}$. On the contrary, as $R$ increases, the Apollonius of Sphere moves closer to the feasible flight region of the UAV. In addition, we have $\mathop {\lim }\limits_{k \to 1} R\left( k \right) = \infty$, $\mathop {\lim }\limits_{k \to 0} R\left( k \right) = 0$, and $\mathop {\lim }\limits_{k \to \infty } R\left( k \right) = 0$. Thus, the $k$ value corresponding to the tangent point is optimal and the tangent point is UAV's optimal location for the next time slot.

As per the geometric properties of the tangency, at the tangent point we have
\begin{equation}\label{mod:c11}
\begin{aligned}
\left(\sqrt{\left(\frac{{kc}}{{{k^2}-1}}\right)^2-H^{2}}+V_{\max}\delta_{t}\right)^2=
\left(x_{n-1}-\frac{k^{2}x_{w}-x_{b}}{k^{2}-1}\right)^2+\left(y_{n-1}-\frac{k^{2}y_{w}-y_{b}}{k^{2}-1}\right)^2,
\end{aligned}
\end{equation}
which simplified as
\begin{equation}\label{mod:c12}
\begin{aligned}
k^{4}(U^{2}+S^{2}+2US)-2k^2(U+SV+2(V_{\max}\delta_{t})^{2}c^2+
d(V+S))+U+V^{2}+2dV=0,
\end{aligned}
\end{equation}
where $U=d^{2}-(V_{\max}\delta_{t})^{2}+H^{2}$, $d=\|\mathbf{q}_{u}[n-1]\|^{2}$, $S=\|\mathbf{q}_{w}\|^{2}-\mathbf{q}_{w}^{T}\mathbf{q}_{u}[n-1]$ and $V=\|\mathbf{q}_{b}\|^{2}-\mathbf{q}_{b}^{T}\mathbf{q}_{u}[n-1]$.
In order to solve \eqref{mod:c11}, we note that
\begin{equation}\label{mod:cc13}
\begin{aligned}
k_{1}^{2}-4k_{0}k_{2}>0,
\end{aligned}
\end{equation}
where $k_{0}=(U+S)^{2}+4(V_{\max}\delta_{t})^{2}H^{2}$, $k_{1}=(2SV-2U^{2}-2US+2UV-4V_{\max}\delta_{t})^{2}(c^2+H^{2}))$, and $k_{2}=(U+V)^{2}+4(V_{\max}\delta_{t})^{2}H^{2}$.
Then, noting $k>1$, the solution to \eqref{mod:c11} of is given by
\begin{equation}\label{mod:c13}
\begin{aligned}
k^{2}=\frac{k_{1}+\sqrt{k_{1}^{2}-4k_{0}k_{2}}}{2k_{0}},
\end{aligned}
\end{equation}
and then the optimal value of $\mathbf{q}_{u}[n]$ can be obtained by solving the following equation
\begin{equation}\label{mod:c17}
\begin{aligned}
\frac{\|\mathbf{q}_{u}[n]-\mathbf{q}_{b}\|^{2}+H^{2}}{\|\mathbf{q}_{u}[n]-\mathbf{q}_{w}\|^{2}+H^{2}}=\frac{k_{1}+\sqrt{k_{1}^{2}-4k_{0}k_{2}}}{2k_{0}}.
\end{aligned}
\end{equation}

For $k<1$, noting $R(k)=\frac{ck}{1-k^{2}}$, the first derivative of $R(k)$ with respect to $k$ is given by
\begin{equation}\label{mod:c8}
\begin{aligned}
(R(k))'=\left(\frac{ck}{1-k^{2}}\right)'=\frac{c(k^{2}+1)}{(k^{2}-1)^{2}}>0,
\end{aligned}
\end{equation}
which indicates that $R(k)$ is a monotonically increasing function of $k$ for $k < 1$. We present the top view of the radius change with respect to $k$ in the case with $k<1$ in Figure.~\ref{fig:11}(d).

From Figure.~\ref{fig:11}(d), we conform that the radius $R(k)$ increases with $k$. This indicates that the Apollonius of Sphere will first gradually be inscribed with the feasible flight region of the UAV and then moves away from this feasible region. Therefore, when feasible flight region of the UAV is inscribed with the Apollonius of Sphere, the value of $k$ is optimal and the corresponding inscribed point is the UAV's optimal position in the next time slot.

Similar to the case with $k>1$, for $k<1$, at the inscribed point we have
\begin{equation}\label{mod:c15}
\begin{aligned}
\left(\sqrt{\left(\frac{{kc}}{{1 - {k^2}}}\right)^2-H^{2}}-V_{\max}\delta_{t}\right)^2=
\left(x_{n-1}-\frac{k^{2}x_{w}-x_{b}}{k^{2}-1}\right)^2+\left(y_{n-1}-\frac{k^{2}y_{w}-y_{b}}{k^{2}-1}\right)^2,
\end{aligned}
\end{equation}
which leads to
\begin{equation}\label{mod:c16}
\begin{aligned}
k^{2}=\frac{k_{1}-\sqrt{k_{1}^{2}-4k_{0}k_{2}}}{2k_{0}}.
\end{aligned}
\end{equation}
Then, the optimal value of $\mathbf{q}_{u}[n]$ can be obtained by solving the following equation
\begin{equation}\label{mod:c17}
\begin{aligned}
\frac{\|\mathbf{q}_{u}[n]-\mathbf{q}_{b}\|^{2}+H^{2}}{\|\mathbf{q}_{u}[n]-\mathbf{q}_{w}\|^{2}+H^{2}}=\frac{k_{1}-\sqrt{k_{1}^{2}-4k_{0}k_{2}}}{2k_{0}}.
\end{aligned}
\end{equation}
This completes the proof of Theorem~\ref{theorem:5}.
\end{IEEEproof}}

\section{Proof of Proposition~1}

\begin{figure}[!t]
\centering
\includegraphics[width=0.6\textwidth]{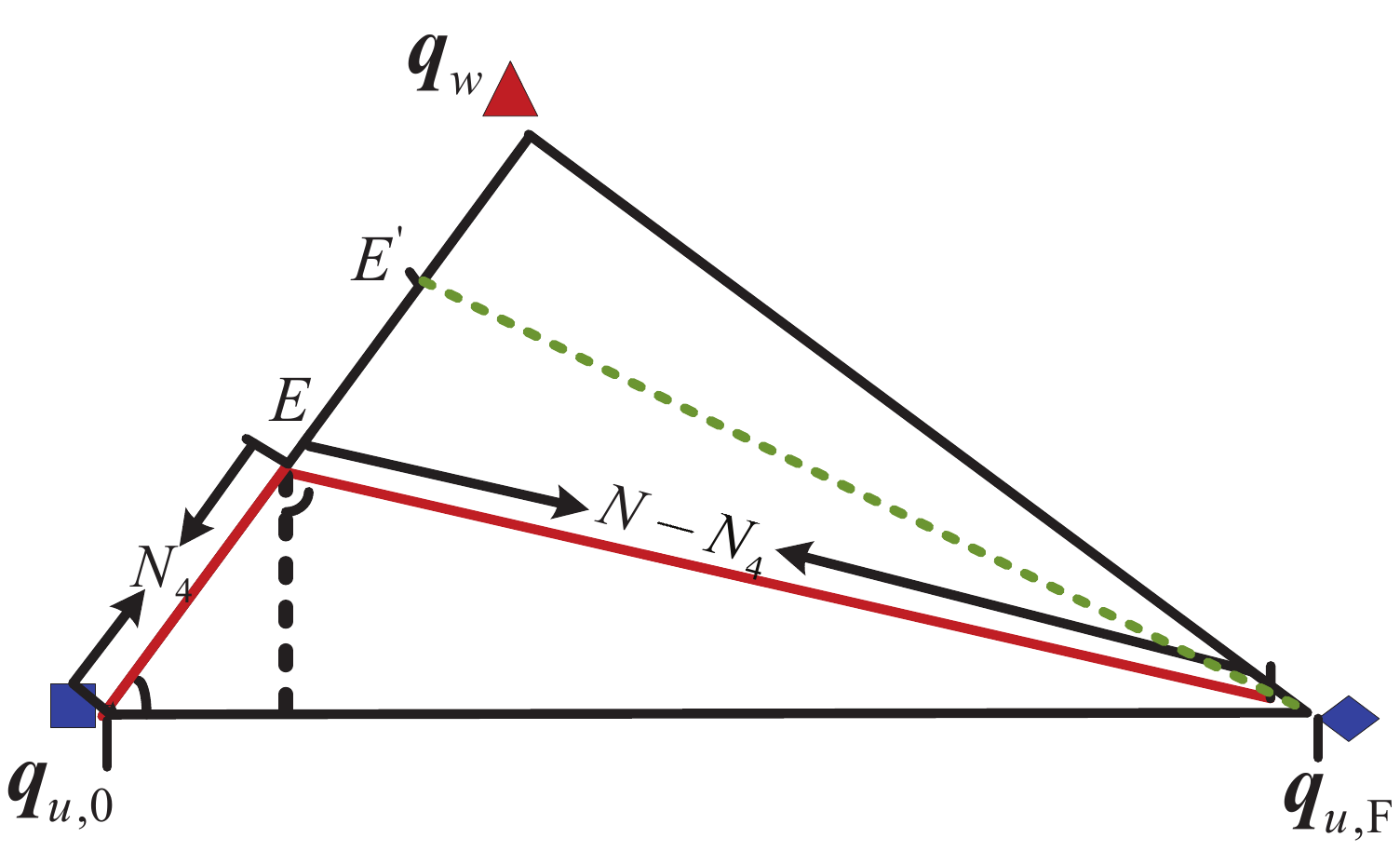}
\vspace{-0.5cm}
\caption{Coordinate position of the intermediate point$\mathbf{q}_{E}$ in the UAV's optimal trajectory.}
\label{fig:99}
\end{figure}

\begin{IEEEproof}
Firstly, we prove the existence of the intermediate point $E$. To this end,
we suppose that the number of time slots required for the UAV to fly from the initial point to point $E$ is $N_{4}$.  Thus, the remaining time slots are $(N-N_{4})$,  as shown
by the red circle in Figure.~\ref{fig:99}. According to the Law of cosines, we have

\begin{align}\label{61}
(N_{4}V_{\max}\delta_{t})^{2}+||{{\mathbf{q}_{u,F}} \!-\! {\mathbf{q}_{u,0}}}||^{2}-2N_{4} V_{\max}\delta_{t}||{{\mathbf{q}_{u,F}} \!-\! {\mathbf{q}_{u,0}}}||\cos (\theta_{1}-\theta_{2})
\!=\!(N-N_{4})^{2}(V_{\max}\delta_{t})^{2}.
\end{align}
After some mathematical calculations, we have
\begin{eqnarray}
\begin{aligned}
N_{4}=\frac{N^{2}(V_{\max}\delta_{t})^{2}-||{{\mathbf{q}_{u,F}} - {\mathbf{q}_{u,0}}}||^{2}}{2V_{\max}\delta_{t}(NV_{\max}\delta_{t}-||{{\mathbf{q}_{u,F}} - {\mathbf{q}_{u,0}}}||\cos(\theta_{1}-\theta_{2}))}.
\end{aligned}
\end{eqnarray}
Due to $\frac{||{{\mathbf{q}_{u,F}} - {\mathbf{q}_{u,0}}}||}{V_{\max}\delta_{t}}<N$, we have
\begin{eqnarray}\label{64}
\begin{aligned}
N_{4}=\frac{N^{2}(V_{\max}\delta_{t})^{2}-||{{\mathbf{q}_{u,F}} - {\mathbf{q}_{u,0}}}||^{2}}{2V_{\max}\delta_{t}(NV_{\max}\delta_{t}-||{{\mathbf{q}_{u,F}} - {\mathbf{q}_{u,0}}}||\cos(\theta_{1}-\theta_{2}))}>0.
\end{aligned}
\end{eqnarray}
Based on \eqref{64}, we know that the point $E$ must exist and the corresponding UAV's trajectory is shown in Figure.~\ref{fig:99}. Considering the constraint that the UAV has to return to the final point by the end of the last time slot and noting the UAV's flight time is an integer, we have
\begin{equation}
\begin{aligned}
 N_{4}=\left\lfloor \frac{N^{2}(V_{\max}\delta_{t})^{2}-||{{\mathbf{q}_{u,F}} - {\mathbf{q}_{u,0}}}||^{2}}{2V_{\max}\delta_{t}(NV_{\max}\delta_{t}-||{{\mathbf{q}_{u,F}} - {\mathbf{q}_{u,0}}}||\cos(\theta_{1}-\theta_{2}))}\right\rfloor,
 \end{aligned}
\end{equation}
Thus, the coordinates of point $E$ is $\mathbf{q}_{E}=N_{4}V_{\max }{\delta _t}\theta_{2}$,
which completes the proof of Proposition~1.
\end{IEEEproof}

\end{document}